\documentclass[prb,twocolumn,amssymb,amsmath,aps,showpacs,superscriptaddress,floatfix,makecell]{revtex4-2}
\usepackage[colorlinks=true,linkcolor=blue,citecolor=blue,urlcolor=blue]{hyperref}
\usepackage{graphicx,bm}
\usepackage{epstopdf}
\usepackage{xcolor}

\begin{document}


\title{Acoustic plasmons in type-I Weyl semimetals}

\date{\today}

\author{A.~N.~Afanasiev}
\email{afanasiev.an@mail.ru} 
\author{A.~A.~Greshnov}
\affiliation{Ioffe Institute, St.Petersburg 194021, Russia}
\affiliation{Moscow Institute of Physics and Technology, Dolgoprudny 141700, Russia}
\author{D.~Svintsov}
\affiliation{Moscow Institute of Physics and Technology, Dolgoprudny 141700, Russia}

\begin{abstract}

Massless chiral fermions emergent in inversion symmetry-breaking Weyl semimetals (WSMs) reside in the vicinity of multiple low symmetry nodes and thus acquire strongly anisotropic dispersion. We investigate the longitudinal electromagnetic modes of  two-component degenerate Weyl plasma relevant to the realistic band structure of type-I WSM. We show that the actual spectrum of three dimensional collective density excitations in TaAs family of WSM is  gaples due to emergence of  acoustic plasmons corresponding to out-of-phase oscillations of the plasma components. These modes exist around the [001] crystallographic direction and are weakly damped, thanks to large difference in the Weyl velocities of the $W_1$ and $W_2$ quasiparticles when propagating along [001]. We show that acoustic plasmons can manifest themselves as slow beatings of electric potential superimposed on fast plasmonic oscillations upon charge relaxation. The revealed acoustic modes can stimulate purely electronic superconductivity, collisionless plasmon instabilities, and formation of Weyl soundarons.

\end{abstract}

\maketitle

\section{Introduction}

The Weyl fermion (WF)-like~\cite{Weyl1929} quasiparticles in the recently discovered {\it Weyl semimetals} (WSMs) manifest themselves in peculiar optical and transport phenomena~\cite{Armitage2018,Burkov2018} related to high energy physics analogies. Imposed by the specific nature of single particle excitations, the plasma modes of WFs in WSM~\cite{DasSarma2009,Sachdeva2015} mimic the ones of the ultrarelativistic plasma~\cite{Silin1960}. The consistent theory of the bulk plasmon dispersion in WSMs taking into account ultaviolet cut-off was elaborated in Refs.\cite{Lv2013_Theory, Hofmann2015_Signature,Thakur2017} for both extrinsic and intrinsic cases, including the case of parallel electric and magnetic fields, related to non-trival topological issue of chiral anomaly inherent to WSM~\cite{Zhou2015}. Due to non-trivial topology, a new type of topological surface collective excitations associated with Fermi arcs arise~\cite{Song2017,Andolina2018,Adinehvand2019}, and the conventional surface plasmon polaritons in time-reversal symmetry breaking WSM become nonreciprocal at zero magnetic field~\cite{Hofmann2016,Chiarello2019}. These features in addition to the characteristic plasmon energy of the order of tens of meV~\cite{Chiarello2019} make WSMs promising materials for THz plasmonics~\cite{Lupi2020}.

In nature, the band structure of real WSM materials is more complicated than given by ''prototypical'' {\it Hamiltonian} due to H.~Weyl~\cite{Weyl1929}, since there is multi-node structure~\cite{Lv2015TaAs,Xu2015TaP,Xu2015TaAs,Xu2015NbAs,Hirayama2015,Ruan2016HgTe,Ruan2016Chalk}. Due to the low local symmetry of the Weyl nodes, WF dispersion acquires strong anisotropy and tilt~\cite{Lee2015,Arnold2016,Hu2016,Klotz2016,Grassano2018}. This leads to qualitative difference in predictions of the {\it prototypical} and {\it realistic} models of WSM, including existence of type-II WSM~\cite{Armitage2018}, novel undamped plasmon modes~\cite{Sadhukhan2020}, photocurrent effects~\cite{Chan2017,Golub2018} and Auger recombination~\cite{Afanasiev2019}.

\begin{figure}[t!]
 \includegraphics[width=0.95\linewidth]{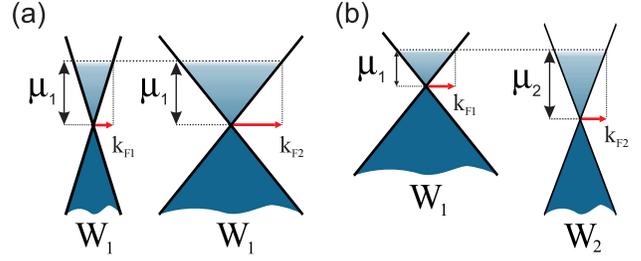}
 \caption{Implementation of the {\em two-component} degenerate plasma in Weyl semimetals: a) homogeneous plasma formed within the single Weyl node group, b) heterogeneous plasma in the multi-group Weyl semimetals (TaAs-like).}
 \label{Fig:Scheme}
\end{figure}

In the realistic band structure of type-I WSMs, there are at least two (non-equivalent) degenerate distributions of carriers resident in the vicinity of different Weyl nodes, which coexist at any direction of the wave vector, thus making the WF plasma in WSM {\it multi-component}. The difference in the Fermi wave vectors $k_F=\mu/v$ (hereinafter $\hbar=1$) between two groups of WFs  can originate solely from the Weyl velocity anisotropy $v_2\neq v_1$ while the Fermi levels are equal $\mu_2=\mu_1$~(Fig.\ref{Fig:Scheme}a) as it happens in the doped WSMs of HgTe family~\cite{HgTe_Fermi_level} possessing single node group, {\it e.g.} in HgTe~\cite{Ruan2016HgTe} under strain or chalcopyrite compounds~\cite{Ruan2016Chalk}. Further we will refer to this case as {\it homogeneous Weyl plasma} (WP). In {\it heterogeneous Weyl plasma}~(Fig.\ref{Fig:Scheme}b) specific to TaAs family~\cite{Lv2015TaAs,Xu2015TaP,Xu2015TaAs,Xu2015NbAs} of WSMs~\cite{TaAs_Not_Semimetal} both unequal values of the chemical potentials and Weyl velocities at the nodes contribute to complexity of the WP composition and properties.

In this paper we show that the structure of plasma excitations in real WSM materials, such as TaAs, NbAs, TaP and NbP, is principally different from the one expected in the ''prototypical'' model~\cite{Lv2013_Theory,Hofmann2015_Signature,Thakur2017}. The dielectric response of two-component WP supports formation of the {\it acoustic plasmons}~\cite{Pines1956}, corresponding to {\it out-of-phase} density oscillations in components, even in isotropic approximation for each component, above some contrast in the WF velocities and/or concentrations (Sec.~\ref{Sec:LongWL}). The anisotropic WP behaves like isotropic one but with the direction dependent WF velocities and effective concentrations (Sec.~\ref{Sec:PlSpec}), so existence of an acoustic mode is dependent on the direction as well. Strongly anisotropic dispersion of WFs and heterogeneous nature of the WP in TaAs family of WSM is favourable for propagation of the acoustic plasmons in some range of directions and the spectrum of plasma excitations in real WSM is {\it gapless}. In order to demonstrate that the acoustic modes can be directly observed in experiments, we calculate the full RPA energy loss function~(Sec.~\ref{Sec:ShortWL}) and investigate~(Sec.~\ref{Sec:Rel}) relaxation of initial perturbation in two-component WP -- the problem related to time domain experimental setups. Emergence of the promising acoustic plasmon related phenomena in WSMs and potential applications are discussed in Sec.~\ref{Sec:Discussion}.

\section{Plasma excitation mode structure of type-I WSM}


\subsection{Acoustic modes in two-component isotropic WP}
\label{Sec:LongWL}

To reveal the gapless nature of WP excitations in WSM we start with the ''minimal'' model describing two-component WP formed by WFs with isotropic Weyl velocities $v_{1,2}$ occupying the Weyl nodes with degeneracy factors $\eta_{1,2}$ and  the Fermi levels $\mu_1$ and $\mu_2$, respectively. The Weyl Hamiltonian for two types of WFs ($j=1,2$) reads
\begin{equation}
	H_{j}=v_j{\bf k} \cdot \boldsymbol{\sigma}.
\end{equation}
The dielectric function of two-component WP is given by
\begin{equation}
	\varepsilon(\omega,q)=1+\Delta\varepsilon_1(\omega,q)+\Delta\varepsilon_2(\omega,q),
	 \label{Eq:TwoCompEpsilon}
\end{equation}
where contributions of the WP components $\Delta\varepsilon_j(\omega,q)$ are additive and are given by the dielectric function of ''prototypical'' WSM~\cite{Lv2013_Theory} with the parameters $v$, $\eta$ and $\mu$, specific to each component.

To find the collective modes of the two-component WP we need to solve the standard equation ${\rm Re}\,\varepsilon(\omega,q)=0$~\cite{Giuliani2005}. In the quasi-classical domain ($\omega, v_j q\ll 2\mu_j$) the contributions to dielectric function from the WP components are given by
\begin{multline}
    \Delta\varepsilon_j(\omega,q)=\frac{q_{_{Tj}}^2}{q^2}\left[1+\frac{\omega}{2 v_j q}\ln\left|\frac{\omega-v_j q}{\omega+v_j q}\right|\right]+\\
    +i\pi\frac{\omega}{2 v_j q}\frac{q_{_{Tj}}^2}{q^2}\Theta(v_j q - \omega),
    \label{Eq:QuasiclassicalEpsilon}
\end{multline}
where $q_{_{Tj}}=\sqrt{2\eta_j\alpha_j/\pi\varkappa_0}k_{_{Fj}}$ is the Thomas-Fermi wavevector, $\alpha_j=e^2/v_j\varkappa_b$ is the Weyl fine structure constant, $\varkappa_b$ is the background dielectric constant, $\varkappa_0=1+\frac{\eta_1\alpha_1}{3\pi}\ln\frac{\omega_{\Lambda}}{2\mu_1}+\frac{\eta_2\alpha_2}{3\pi}\ln\frac{\omega_{\Lambda}}{2\mu_2}$ describes the renormalization of $\varkappa_b$~\cite{Lv2013_Theory,Hofmann2015_Signature}, and $\omega_{\Lambda}$ is the cut-off energy of the WF model~\cite{Hofmann2015_Signature,AbrikosovBeneslavskii1971_Cutoff}.

\begin{figure}[t]
 \includegraphics[width=0.95\linewidth]{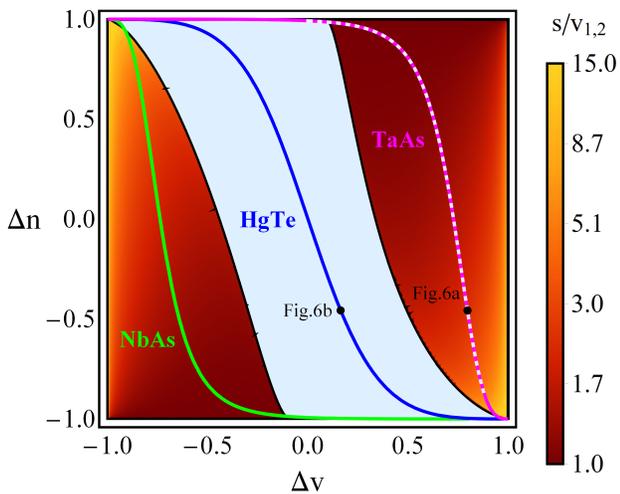}
 \caption{Diagram of the two-component WP compositions supporting the acoustic modes and their velocities $s$ according to Eq.~(\ref{Eq:Ac_Velocity}). Solid lines correspond to homogeneous plasma (\textit{e.g.} doped HgTe~\cite{Ruan2016HgTe} under strain or chalcopyrites compounds~\cite{Ruan2016Chalk}) and heterogeneous plasma of TaAs ($\mu_2/\mu_1=5$~\cite{Lv2015TaAs}) and NbAs ($\mu_1/\mu_2=8.25$~\cite{Xu2015NbAs,Grassano2018}). Dashed part of the TaAs composition line corresponds to the anisotropic WP in (110) and ${\rm (1\bar{1}0)}$ planes within the numerically predicted band structure~\cite{Lee2015,Grassano2018}.}
 \label{Fig:PhaseDiagram}
\end{figure}

We focus on the emergence of {\it acoustic plasmon} modes $\omega_{ac}(q)=s q$~\cite{Pines1956} lying in-between the WF dispersion lines of the slow and fast plasma components $v_1 q<\omega<v_2 q$ (or vice versa). In the limit of $q\rightarrow 0$, ${\rm Re}\,\varepsilon(\omega_{ac}(q), q)=0$ transforms into the equation for "sound" velocity
\begin{equation}
    \label{Eq:Ac_Velocity}
    \left|\frac{s-v_1}{s+v_1}\right|^{s q_{_{T1}}^{2}/2 v_1}\left|\frac{s-v_2}{s+v_2}\right|^{s q_{_{T2}}^{2}/2v_2}={\rm e}^{-(q_{_{T1}}^{2}+q_{_{T2}}^{2})}
\end{equation}
dependent on the ratios of WF velocities $\delta v=v_2/v_1$ and Thomas-Fermi wave vectors $\delta q_{_T}=q_{_{T2}}/q_{_{T1}}$. Using the relation
\begin{equation}
    \frac{q_{_{T2}}}{q_{_{T1}}}=\left(\frac{\eta_2 n_2^2 v_1^3}{\eta_1 n_1^2 v_2^3}\right)^{1/6}
\end{equation}
it is convenient to consider solution of Eq.~(\ref{Eq:Ac_Velocity}) in terms of the dimensionless parameters
\begin{gather}
    \Delta v=\frac{E_2(q)-E_1(q)}{E_2(q)+E_1(q)}=\frac{v_2-v_1}{v_2+v_1}
    \label{Eq:QuantumDisbalance}\\
    \Delta n=\frac{n_2-n_1}{n_2+n_1}
	\label{Eq:StatDisbalance}
\end{gather}
expressing degree of the quantum mechanical and statistical disbalances, respectively, between the WP components. Here $E_j(q)=v_{j} q$ denotes the dispersion of WFs and $n_j$ are concentrations in the plasma components
\begin{equation}
    n_j=\frac{\eta_j}{6 \pi^2}\frac{\mu_j^3}{v_j^3}.
\end{equation}
We present existence of the acoustic plasmon solution and its velocity, in the units of the smaller WF velocity -- $v_1$ at $\Delta v>0$ and $v_2$ at $\Delta v<0$, by color diagram in Fig.~\ref{Fig:PhaseDiagram}. It is asymmetric with respect to permutation of plasma components ({\textit i.e.} to the simultaneous transformations $\Delta n \rightarrow -\Delta n$ and $\Delta v \rightarrow -\Delta v$) representing the effect of the difference in Weyl node degeneracy on the acoustic plasmon formation. Namely, we consider $\delta\eta=\eta_2/\eta_1=2$ characteristic for TaAs family of WSMs, so at $\Delta v>0$ the number of the fast nodes is greater then the slow ones, while at $\Delta v<0$ the situation is opposite.


Since WF distribution implies the fixed Fermi level ratio $\delta\mu=\mu_2/\mu_1$ but not the concentrations $n_j$, realistic two-component WP compositions are presented on the diagram by the curves
\begin{equation}
    \label{Eq:CompositionLine}
    \Delta n(\Delta v)=\frac{\delta \eta \delta \mu^3 (1-\Delta v)^3-(1+\Delta v)^3}{\delta \eta \delta \mu^3 (1-\Delta v)^3+(1+\Delta v)^3}
\end{equation}
at the specific $\delta\mu$ and $\delta \eta$. Fig.~\ref{Fig:PhaseDiagram} shows that in the {\it homogeneous} plasma (depicted by the blue line) acoustic mode arises only at extreme difference between velocities of the WP components $\Delta v \gtrsim 0.99$ i.e. $v_2/v_1\gtrsim 20$, when the $\Delta n(\Delta v)$ line~(\ref{Eq:CompositionLine}) at $\delta \mu =1$ and $\delta \eta=1$ enters the acoustic plasmon domain. With the increase of heterogeneity $\delta \mu$, WP composition line crosses the border at moderate $\Delta v$ and the acoustic mode tends to exist at wider range of $\Delta v$. Namely, the {\it heterogeneous} WP lines corresponding to TaAs and NbAs lie well inside the acoustic plasmon domain. The upper curves lying predominantly at $\Delta n> -\Delta v$ correspond to $\mu_2>\mu_1$, the lower ones - to $\mu_2 < \mu_1$. 

The similar mechanism governs the formation of acoustic spin plasmon in spin-polarized degenerate two-dimensional electron gas~\cite{Agarwal2014} at $P=(n_{\uparrow}-n_{\downarrow})/(n_{\uparrow}+n_{\downarrow})>1/7$. In this case quasiparticles with different spins posses equal parabolic dispersions and thus Fermi velocities $v_{_{Fj}}$ are determined by the concentrations of plasma components. Therefore, the only independent parameter controlling the emergence of linearly dispersing collective modes is spin polarization $P$ analogous to $\Delta n$  in~(\ref{Eq:StatDisbalance}).

\begin{figure}[t]
 \includegraphics[width=0.95\linewidth]{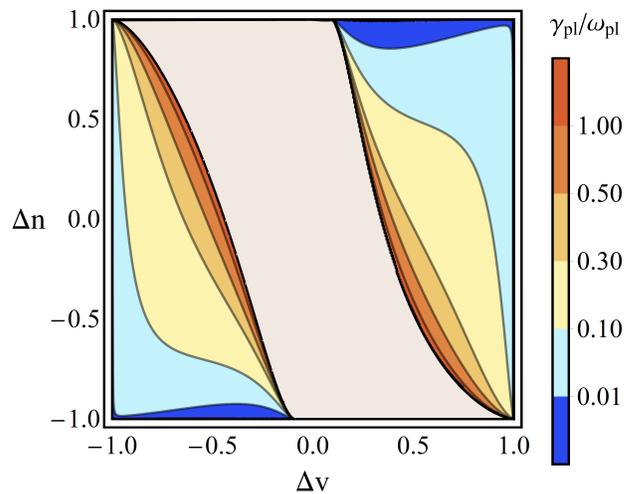}
 \caption{Contour plot of the dimensionless damping parameter~(\ref{Damping}) for acoustic plasmon in the region of its existence in WSM with $\eta_2/\eta_1=2$.}
 \label{Fig:Damping}
\end{figure}

Acoustic plasmon mode lies in the domain of small, but nonzero Landau damping provided by the interband single particle excitations in the fast component. Therefore, the criteria of its existence is smallness of the collisionless damping $\gamma_{ac}(q)=\left.{\rm Im}\varepsilon/\frac{\partial{\rm Re}\,\varepsilon}{\partial\omega}\right|_{\omega_{ac}(q)}$ compared to the plasmon frequency $\omega_{ac}(q)$. In the quasi-classical domain, the ratio $\gamma_{ac}(q)/\omega_{ac}(q)$ is constant and its explicit form is given by
\begin{equation}
    \label{Damping}
    \frac{\gamma_{ac}(q)}{\omega_{ac}(q)}=\frac{\pi}{2}\frac{s}{v_2}\frac{q_{_{T2}}^2(s^2-v_1^2)(v_2^2-s^2)}{q_{_{T1}}^2 v_1^2 (v_2^2-s^2)-q_{_{T2}}^2 v_2^2 (s^2-v_1^2)}.
\end{equation}
Fig.~\ref{Fig:Damping} displays the contour plot of $\gamma_{ac}(q)/\omega_{ac}(q)$ for two-component WP compositions supporting the acoustic collective mode~(Fig.~\ref{Fig:PhaseDiagram}). With account for damping, the domain of acoustic plasmon existence shrinks to the area of greater $|\Delta v|$, but TaAs and NbAs lines reside in the region of well-defined acoustic mode. 

\subsection{Plasmon spectrum of type-I WSM}
\label{Sec:PlSpec}
\begin{figure}[t]
 \includegraphics[width=0.95\linewidth]{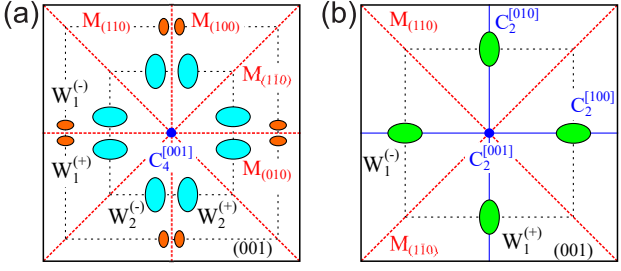}
 \caption{Projection of the band structure of a) TaAs family WSMs and b) HgTe family WSMs on the (001) crystallographic plane. Filled ellipses show the cross sections of the approximate Fermi surfaces. Red dashed lines denote mirror symmetry planes perpendicular to (001). Blue lines and dots correspond to symmetry axes.
\label{Fig:BandStruct}}
\end{figure}

In experimentally relevant WSMs, WF reside in the vicinity of multiple Weyl nodes belonging to the single or multiple node groups $W_n$.  Weyl nodes of the same group are linked by the crystal symmetry operations $g_{i}\in \mathcal{G} \times \mathcal{T}$ where $\mathcal{T}$ is the time reversal and $\mathcal{G}$ is the point group of WSM, e.g. $C_{4v}$ for TaAs family WSM~\cite{Armitage2018} and $D_{2d}$ for HgTe family WSMs~\cite{Ruan2016HgTe,Ruan2016Chalk}, see Fig.~\ref{Fig:BandStruct}. Due to the low local symmetry, WF states near the node $W_{n,i}$ are described by the generalized Weyl Hamiltonian
\begin{equation}
	\label{Eq:HrealWSM}
	H_{i,n}={\bf v}_{n}^{(t)}g_i {\bf k} \sigma_0 + \hat{v}_{n} g_i {\bf k} \cdot \boldsymbol{\sigma}
\end{equation}
where ${\bf k}$ is the WF wave vector with respect to the position of $W_{n,i}$ in the Brillouin zone, $\sigma_0$ is the identity matrix, ${\bf v}_{n}^{(t)}$ and $\hat{v}_n$ are tilt and Weyl velocities respectively.

The dielectric response of the multi-node WF system is given by the sum of individual contributions from each Weyl node. Linearity of the Hamiltonian~(\ref{Eq:HrealWSM}) in ${\bf k}$ allows one to relate the contribution to the dielectric function from a node $W_{n,i}$ with that of the ''prototypical'' WSM with the Fermi level $\mu_n$ and the ''average'' Weyl velocity  $v_n=|{\rm det}\,\hat{v}_n|^{1/3}$, $\Delta\varepsilon_n(\omega,q)$, via a linear coordinate transform~\cite{Afanasiev2019}. Thus, the total $\varepsilon(\omega,{\bf q})$ is given by
\begin{equation}
    \label{Eq:EpsilonWSM}
    \varepsilon(\omega,{\bf q})=1+\sum\limits_{n,i}\Delta\varepsilon_{n}(\omega-{\bf v}_{n}^{(t)}g_i{\bf q},\left|v_n^{-1}\hat{v}_{n}g_{i}{\bf q}\right|).
 \end{equation}
This expression is applicable to the arbitrary type-I WSM until tilt is large enough to form the type-II WSM. Since every node has a time reversal partner with the same chirality but opposite tilt velocity, the expansion of~(\ref{Eq:EpsilonWSM}) in powers of $|{\bf v}_n^{(t)}|/|\hat{v}_n \bf{e_q}|$ starts with the quadratic terms and the tilt-induced Doppler shift in $\varepsilon(\omega,{\bf q})$ can be reasonably neglected at typical values of tilt in experimentally available WSM~\cite{Grassano2018}.

Following properties of the individual WFs, WP in type-I WSM is substantially anisotropic and multi-component because of non-equivalence in energies of the WFs from different nodes (of either the same or different groups) at a given wave vector $\bf q$. In the directions and planes of high symmetry the WP of experimentally available WSMs reduces to two components, and the dielectric response is equivalent to the one of isotropic two-component WP considered in Sec.~\ref{Sec:LongWL}, but with the direction-dependent parameters
\begin{gather}
n_j \rightarrow n_{j}({\bf e_q})=\eta_j [k_{_{Fj}}({\bf e_q})]^3/6 \pi^2\\
v_j \rightarrow v_j({\bf e_q})=\hat{v}_j {\bf e_q},
\end{gather}
where ${\bf e_q}={\bf q}/q$. Thus, $\Delta n({\bf e_q})$ and $\Delta v({\bf e_q})$ in (\ref{Eq:StatDisbalance}), (\ref{Eq:QuantumDisbalance}) and (\ref{Eq:CompositionLine}) become ${\bf e_q}$-dependent and position of the corresponding WP composition line and the range of actual $\Delta v({\bf e_q})$ (\textit{i.e.} points on it) is determined by specific band structure parameters of WSM: heterogeneity $\delta \mu$, the values and anisotropy of Weyl velocities $v_j({\bf e_q})$, and orientation of the principal axes of $\hat{v}_n$ with respect to crystallography. Below we consider spectrum of the collective excitations of WP in WSMs of TaAs and HgTe families, within the realistic band structure.

Weyl nodes present in TaAs-like WSMs are generally located in low-symmetry points of the Brillouin zone close to high-symmerty mirror planes (100) and (010)~\cite{Grassano2018,Lee2015,Lv2015TaAs,Xu2015TaAs,Xu2015NbAs}. The overall 24 nodes are divided into two groups with different Fermi levels: 8 nodes of type $W_1$ lying in (001) plane, and 16 nodes $W_2$ reside in the two planes perpendicular to [001] with $|k_z|\neq 0$. According to numerical calculations of the band structure~\cite{Grassano2018,Lee2015}, principal axes of Weyl velocity tensors $\hat{v}_{1,2}$ for both groups $W_{1,2}$ are almost collinear to the cubic ones. Since the Weyl nodes of opposite chiralities are connected via the mirror reflections and time reversal, we can neglect this small misorientation by the same reasons as it was done with the contribution of tilt velocity to~(\ref{Eq:EpsilonWSM}). We also ignore the minor difference between $v_{x,y}$ for $W_{1,2}$ and consider the averaged $\bar{v}_{x,y}=v_{x,y}^{(1)}/2+v_{x,y}^{(2)}/2$ instead. In this section $v_{c}^{(n)}$, $c=x,y,z$ denote the principal values of the Weyl velocity tensor $\hat{v}_n$ in group $W_n$.

\begin{figure}[t]
 \includegraphics[width=0.95\linewidth]{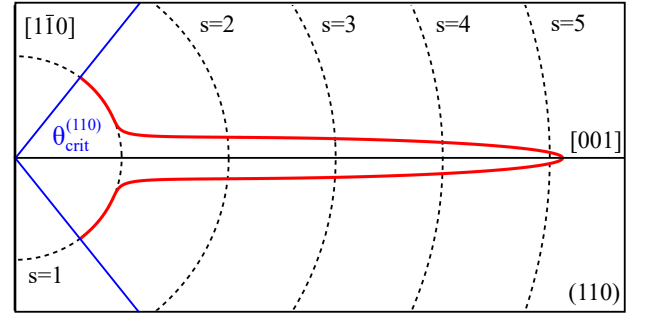}
 \caption{Anisotropy (red curve) of acoustic plasmon velocity $s(\theta)$ (in units of $v_{(110)}^{(1)}(\theta)$) in the upper half of the (110) crystallographic plane of TaAs with Weyl velocities taken from~\cite{Grassano2018}. Directions in which acoustic plasmon can propagate lie inside the $2 \theta_{\rm crit}^{(110)}$ angle formed by the two solid blue lines. 
 \label{Fig:AcVelAnisotropy}}
\end{figure}

With account for the numerically predicted band structure of TaAs-like WSMs shown in Fig.~\ref{Fig:BandStruct}a, the embodied WP is exactly two-component only in (110), ${\rm (1\bar{1}0)}$ and (001) planes. Namely, the dielectric response in the (001) plane is given by the WFs belonging to the two sets of Weyl nodes connected via $C_4$ rotation around [001] axis. Due to the multigroup structure of TaAs-like WSMs, WFs with $v_{(001)}({\bf e_q})$ and $v_{(001)}(C_4{\bf e_q})$ can belong to the both degenerate distributions described by the Fermi levels $\mu_{1,2}$ respectively. Thus WP in the (001) plane is effectively homogeneous with the Fermi energy $\sqrt{(\eta_1\mu_1^2+\eta_2\mu_2^2)/(\eta_1+\eta_2)}$. The maximal velocity difference between the plasma components is achieved in [100] and [010] directions and its magnitude $\bar{v}_x/\bar{v}_y=1.56$ guaranties (see Sec.~\ref{Sec:LongWL}) the absence of acoustic plasmons in the (001) plane. 

In the (110) crystallographic plane, WFs belonging to the $W_1$ and $W_2$ groups of Weyl nodes form, respectively, the slow and the fast components of anisotropic heterogeneous WP. According to the phase diagram shown in Fig.~\ref{Fig:PhaseDiagram}, the part of the TaAs composition line which describes the WP in the (110) plane (denoted by the white dashing) lies well inside the acoustic plasmon domain and the linearly dispersive collective mode is supported. The maximal speed of sound is achieved in the most favorable direction [001] due to the giant inequality~\cite{Lee2015,Lv2015TaAs,Xu2015TaAs,Xu2015NbAs} of $v_z^{(2)}/v_z^{(1)}\approx 15$~\cite{Grassano2018} between node groups. This direction is presented on the TaAs composition line by the right endpoint of the dashed part at $\Delta v_{[001]}=0.875$. With the increase of the polar angle $\theta$ with respect to the [001] direction, the difference in the velocities of plasma components   
\begin{equation}
v_{(110)}^{(j)}(\theta)=\sqrt{v_{[110]}^2\sin^2 \theta + \left[v_{z}^{(j)}\right]^2\cos^2 \theta}
\end{equation} 
becomes smaller and the speed of acoustic plasmon rapidly decreases~(see Fig.~\ref{Fig:AcVelAnisotropy}). The point $\{ \Delta n\left(\Delta v_{(110)}(\theta)\right),\Delta v_{(110)}(\theta)\}$ describing the (110) WP on the phase diagram moves, respectively, along the TaAs composition line towards $\Delta v_{[110]}=0$ corresponding to [110] (or ${\rm [1\bar{1}0]}$) direction, when $v_{(110)}^{(1)}(\theta)$ and $v_{(110)}^{(2)}(\theta)$ merge into $v_{[110]}^2=(\bar{v}_x^2+\bar{v}_y^2)/2$.  Thus, acoustic plasmon does not propagate in the range of directions in the (110) plane lying between [110] and the critical direction corresponding to the polar angle $\theta^{(110)}_{\rm crit}(\delta \mu)\approx 0.9$.

Anisotropy of the acoustic plasmon speed in the (110) plane is also affected by the crystal point symmetry. Due to the presence of the $M_{(1\bar{1}0)}$ mirror plane, $s({\bf e_q})$ plot presented in Fig.~\ref{Fig:AcVelAnisotropy} is symmetric with respect to reversion of the vertical axis, i.e. to the $[1\bar{1}0]\leftrightarrow [\bar{1}10]$ permutation. However, acoustic plasmons which propagate towards [100] ($0<\theta<\pi/2$) and opposite to it ($\pi/2<\theta<\pi$) are not equivalent in inversion symmetry-breaking WSMs like TaAs. The lack of inversion symmetry is introduced into the dielectric function of type-I WSM~(\ref{Eq:EpsilonWSM}) via nonzero but rather small tilt velocity. Therefore, the non equivalence of acoustic plasmon speed $s(\theta)\neq s(\pi-\theta)$ arise as a second order effect [see discussion after Eq.~(\ref{Eq:EpsilonWSM})] in $v_{z,2}^{(t)}/v_z^{(2)}\ll 1$ ($v_{z,1}^{(t)}=0$ for TaAs family of WSMs) and the $s(\theta)$ plot for $\theta>\pi/2$ is qualitatively the same as Fig.~\ref{Fig:AcVelAnisotropy} signifying the propagation of acoustic plasmons in a range of directions near $[00\bar{1}]$.

In an arbitrary plane $(kl0)$ WF dispersion in TaAs-like WSMs is described by the two pairs of Weyl velocities: $v_z^{(1,2)}$ in the [001] direction and $v_{(001)}({\bf e_q})$ and $v_{(001)}(C_4{\bf e_q})$ perpendicular to it. Therefore, the embodied WP is four-component in $(kl0)$ and the velocity contrast induced by the giant inequality of $v_z^{(2)}$ and $v_z^{(1)}$ is lowered at a given angle $\theta$ in comparison to the (110) plane, so that $\Delta v_{(kl0)}(\theta) < \Delta v_{(110)}(\theta)$. With deviation from the (110) plane, acoustic plasmon propagate in a narrower range of directions near $[001]$ (and $[00\bar{1}]$) characterized by smaller critical angle $\theta_{\rm crit}^{(kl0)} < \theta_{\rm crit}^{(110)}$. We expect $s({\bf e_q})$ in $(kl0)$ to become even more anisotropic with respect to [001], and the $s({\bf e_q})$ plot shown in Fig.~\ref{Fig:AcVelAnisotropy} to squeeze in the vertical direction. Thus, the collective mode spectrum of TaAs-like WSMs is {\it gapless} due to formation of weakly damped acoustic plasmon with strongly anisotropic dispersion.

In the case of HgTe family WSMs~\cite{Ruan2016HgTe,Ruan2016Chalk} of $D_{2d}$ point group, eight Weyl nodes of the single group $W_1$ lie exactly in planes (100) and (010). The directions of principal axes of $\hat{v}$ have not been studied numerically yet and symmetry considerations dictate that one of them should be practically perpendicular to the vertical plane to which the particular Weyl node belong. For the sake of simplicity, we assume that the principal axes of $\hat{v}$ coincide with the cubic axes, see Fig.~\ref{Fig:BandStruct}b. WF plasma is precisely two-component (except for the (110) set of planes and [001] direction), being formed by the quasiparticles resident in the vicinity of the Weyl nodes linked by the mirror symmetry operations $M_{(110)}$ and $M_{(1\bar{1}0)}$. However, the velocity anisotropy is not strong enough~\cite{Ruan2016HgTe,Ruan2016Chalk} for this homogeneous plasma to support the acoustic plasmon even in the most favorable directions [100] and [010], where $\Delta v$ is maximal. Therefore, the collective plasma mode spectrum of HgTe family WSMs is {\it gapped}, being dominated solely by the {\it optical plasmon mode}, corresponding to the {\it in-phase} density oscillations in the plasma components. We point out that at arbitrary orientation of the principal axes of $\hat{v}$ homogeneous plasma of the HgTe family WSMs become multi-component and less favorable for the formation of acoustic modes by the same reason as discussed for TaAs $(kl0)$ planes, and the predicted plasmon structure remains unchanged.

In the heterogeneous WP with $\delta \mu \approx 1$, $\omega_{opt}$ stays well below the edge of the single particle continuum given by $\omega=2\mu_1$, where $\mu_1$ denote the smaller Fermi level. In this domain $\Delta\varepsilon_{j}(\omega,0)=-\omega_{j}^2/\omega^2$ and the optical plasmon frequency is given by the conventional expression $\omega_{opt}=\sqrt{\omega_{1}^2+\omega_{2}^2}$, where $\omega_{j}\sim \sqrt[3]{n_j}$ is the plasmon frequency of the isolated WP components. As the heterogeneity $\delta \mu$ increases, $\omega_{opt}$ approaches $2\mu_1$ and the singularity in $\rm{Re}\varepsilon(\omega,0)$ (at $\omega=2\mu_1$) inherent to the degenerate plasma with linear dispersion~\cite{Falkovsky2011} affects the optical plasmon energy~\cite{Zhou2015}. At $T=0$ it prevents $\omega_{opt}$ from entering the single particle continuum domain, while at finite temperature the singularity is smeared and the optical plasmon vanish when $2\mu_1\approx T$.

\subsection{Full RPA energy loss function and short-wavelength plasmon dispersion}
\label{Sec:ShortWL}

\begin{figure}[t]
 \includegraphics[width=0.95\linewidth]{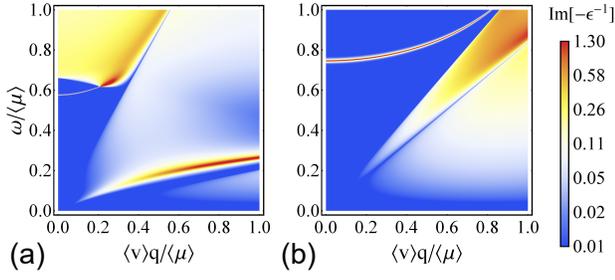}
 \caption{Full RPA energy loss function for the cases of (a) heterogeneous WP in TaAs supporting the acoustic plasmon mode and (b) homogeneous plasma of HgTe-like WSM (see Fig.~\ref{Fig:PhaseDiagram}). Here $\langle \mu \rangle =(\mu_1+\mu_2)/2$ and $\langle v \rangle=(v_1+v_2)/2$.}
 \label{Fig:LossFunction}
\end{figure}

Full RPA energy loss function calculations, shown in Fig.~\ref{Fig:LossFunction}, confirm the plasmon structure of type-I WSMs predicted in the previous section. Particularly, in the case of heterogeneous plasma of TaAs with the composition shown in Fig.~\ref{Fig:PhaseDiagram},  ${\rm Im}[-\varepsilon^{-1}(\omega,{\bf q})]$ spectrum (at various $q$) demonstrates (Fig.~\ref{Fig:LossFunction}a) the {\it two} distinct peaks corresponding to optical and acoustic collective modes, while in homogeneous case (Fig.~\ref{Fig:LossFunction}b) there is {\it only one} optical plasmon peak. Consequently, acoustic plasmon dispersion in WSMs should be directly measurable via the energy loss spectroscopy experiments except for the domain of extremely low $q$, where ${\rm Im}[-\varepsilon^{-1}(\omega_{ac}({\bf q}),{\bf q})]=[{\rm Im}\varepsilon(\omega_{ac}({\bf q}),{\bf q})]^{-1}\sim q^2$, but the acoustic plasmon is well-defined according to Fig.~\ref{Fig:Damping}.

Plasmon dispersion in the short wavelength domain is generally determined by the full RPA dielectric function~\cite{Lv2013_Theory}. According to Fig.~\ref{Fig:LossFunction}a, acoustic plasmon stays well-defined in this region, its dispersion acquires small non-linearity and the optical and acoustic modes tend to merge with WF's dispersion lines $\omega=v_{2,1}({\bf e_q})q$, respectively. 

In the limit $v_2/v_1 \gg 1$, dispersion and damping of acoustic and optical modes can be studied analytically. When the frequency $\omega$ approaches the dispersion line $v_F q$, dielectric response of the degenerate gas with linear dispersion is determined by the singularity in ${\rm Re}\,\varepsilon(\omega,{\bf q})$, being logarithmic in WSMs~\cite{Lv2013_Theory} and square-root in graphene~\cite{Hwang2007}. Since $\omega_{ac}({\bf q})\approx v_1({\bf e_q})q\ll v_2({\bf e_q})q$, 
\begin{gather}
\Delta\,\varepsilon_{1}(\omega,{\bf q})=\frac{\eta_1\alpha_1}{6\pi}\left(\frac{6 k_{_{F1}}^2}{q^2}-1\right)\ln\left|1-\frac{\omega}{v_1 q}\right|\\
\Delta\, \varepsilon_{2}(\omega,{\bf q})=\frac{q^2_{_{T2}}}{q^2}+i\pi\frac{\omega}{2 v_2 q}\frac{q_{_{T2}}^2}{q^2}
\end{gather}
and the corresponding acoustic plasmon dispersion and damping
\begin{gather}
\omega_{ac}({\bf q})=v_1 q \left[1+\exp\left(-2\frac{\delta q_{_T}^2}{1-q^2/6 k_{_{F1}}^2}\right)\right]\\
\frac{\gamma_{ac}}{\omega_{ac}}=6\pi \frac{\delta q_{_T}^2}{\delta v}\frac{1}{1-q^2/6 k_{_{F1}}^2}\exp\left(-2\frac{\delta q_{_T}^2}{1-q^2/6 k_{_{F1}}^2}\right).
\end{gather}
Due to the pronounced logarithmic singularity, $\omega_{ac}({\bf q})$ has an endpoint at $|{\bf q}|=\sqrt{6}k_{_{F1}}({\bf e_q})$. The nonlinear correction to $\omega_{ac}({\bf q})$ and the damping are small, independently on the magnitude of $\delta q_{_T}$. The similar singularity-induced endpoint emerge in the the optical plasmon dispersion. It was previously obtained from numerical calculations in one-component WP within full RPA approximation~\cite{Lv2013_Theory,Zhou2015}.

\begin{figure}[t]
 \includegraphics[width=0.95\linewidth]{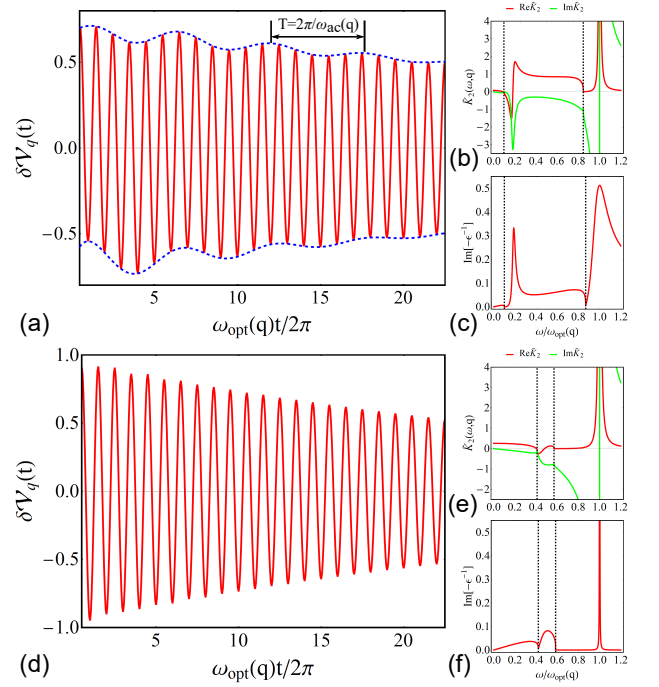}
 \caption{Relaxation dynamics (a),(d) of the self-consistent potential $\delta\mathcal{V}_{q}(t)$~(\ref{Eq:RelDyn}), relaxation spectrum $\tilde{K}_2(\omega,q)$ (b),(e) after~(\ref{Eq:RelPropagator}) and full RPA loss function (c),(f) at $\langle v \rangle q=0.4 \langle \mu \rangle$ in heterogeneous (Fig.~\ref{Fig:LossFunction}a) and homogeneous (Fig.~\ref{Fig:LossFunction}b) WP respectively. Vertical dashed black lines correspond to the boundaries $\omega=v_{1,2}q$ of intraband and interband single particle excitations.}
 \label{Fig:Relaxation}
\end{figure}

\section{Relaxation of initial perturbation in two-component WP}
\label{Sec:Rel}
 
To demonstrate experimental feasibility of acoustic plasmons in WSMs, we consider the relaxation of initial perturbation in isotropic two-component WP. This problem underlies the direct observation of plasma oscillations in time resolved measurements of voltage between injector and detector contacts placed on a sample~\cite{Ashoori1992,Kumada2014}. Following the method described in~\cite{Landau_V10} we study the self-consistent response to the weak initial plane wave-like perturbation of the fast component within the set of linearized collisionless (since the effective dimensionless interaction strength is small $\tilde{\alpha}_j=e^2/\hbar v_j \varkappa_b \varkappa_0 \ll 1$) Boltzman equations for both components (see Appendix~\ref{Sec_AP:Rel}). The time evolution of the dimensionless self consistent potential $\delta \mathcal{V}_{q}(t)$ per one perturbed WF is given by the Fourier transform of the dimensionless RPA relaxation propagator~\cite{Giuliani2005,Badalyan2013} of the fast component $\tilde{K}_2(\omega,q)$
\begin{gather}
    \label{Eq:RelDyn}
    \delta\mathcal{V}_{q}(t)=\int\limits_{-\infty}^{+\infty}\frac{d\omega}{2\pi}{\rm e}^{-i \omega t} \tilde{K}_2(\omega,q),\\
    \label{Eq:RelPropagator}
     \tilde{K}_2(\omega,q)=\frac{1}{\varepsilon(\omega,q)}\frac{\tilde{\Pi}_2(0,q)-\tilde{\Pi}_2(\omega,q)}{i\omega}.
\end{gather}
Here $\tilde{\Pi}_2(\omega,q)$ is the dimensionless quasicalssical non-interacting polarizability~(\ref{Eq_Ap:NonIntPolarizability}) and $\varepsilon(\omega,q)$ is given by Eqs.~(\ref{Eq:TwoCompEpsilon}) and~(\ref{Eq:QuasiclassicalEpsilon}). The distant stages of the relaxation process are determined by the poles of the relaxation propagator~(\ref{Eq:RelPropagator}) given by $\varepsilon(\omega,q)=0$ and the asymptotic behavior of $\delta \mathcal{V}_{q}(t)$ has the form~\cite{Landau_V10}
\begin{equation}
    \delta\mathcal{V}_{q}(t)\sim\sum\limits_{l}{\rm e}^{-i\,\omega_{l}({q})t}{\rm e}^{-\gamma_{l} t},
\end{equation}
where $l$ denotes the plasmon dispersion branch.

To investigate the number and damping of the complex zeros of the full dielectric function given by Eqs.~(\ref{Eq:TwoCompEpsilon}) and~(\ref{Eq:QuasiclassicalEpsilon}), we calculate the relaxation dynamics of $\delta \mathcal{V}_{q}(t)$ by performing the one-dimensional numerical integration in~(\ref{Eq:RelDyn}) over real frequencies. In agreement with the collective mode structure described in Sec.~\ref{Sec:LongWL}, relaxation spectrum $\tilde{K}_2(\omega,q)$ of TaAs~(Fig.~\ref{Fig:Relaxation}b) is dominated by two signatures corresponding to acoustic and optical plasmon peaks in the full RPA energy loss function~(Fig.~\ref{Fig:Relaxation}c). As a result, high-frequency oscillations of $\delta \mathcal{V}_{q}(t)$ relevant to optical plasmon (which has entered the Landau damping region at the considered wave vector) are affected by the distinct low-frequency modulation at $\omega_{ac}(q)$ with lower lifetime, demonstrating the direct experimental accessibility of acoustic plasmons in TaAs-like WSMs. According to Fig.~\ref{Fig:Relaxation}d and Fig.~\ref{Fig:Relaxation}e relaxation of initial perturbation in homogeneous WP in governed by the only present optical mode (Fig.~\ref{Fig:Relaxation}f), as expected. Since optical plasmon is undamped in this case~(Fig.~\ref{Fig:LossFunction}), we have introduced an artificial collisional damping ${\rm Im}\,\varepsilon=\,10^{-2}$ to regularize the $\omega$ integration in~(\ref{Eq:RelDyn}).

\section{Discussion and conclusions}
\label{Sec:Discussion}

In real TaAs-family WSMs, Weyl nodes of the same group with opposite chiralities are located close to each other with respect to the size of the first Brillouin zone. The typical values of this internodal distance in TaAs is $|\textbf{Q}|\approx 0.05 \frac{\pi}{a}$~\cite{Lv2015TaAs}, where $a$ is the lattice constant. Therefore, internode transitions may give significant contribution to the total dielectric function and thus affect the plasmon dispersion as it happens with the emergence of additional harmonics in the Friedel oscillation pattern~\cite{Lv2013_Theory}. In Appendix~\ref{Sec_Ap:Chirality} we show that such chirality-flip transitions lead to additional (and rather small) renormalization of the background dielectric constant provided by the resolvability condition of the Weyl nodes $k_F^{(n)}({\bf e_{_{Q_n}}})\ll |{\bf Q_n}|$ belonging to the group $W_n$. Thus, the revealed collective mode structure remains qualitatively unchanged.

We believe that our results will stimulate further research on the acoustic plasmon-mediated phenomena in WSMs. In particular, TaAs-family WSMs may demonstrate the similar unconventional superconductivity as twisted bilayer graphene~\cite{Cao2018}, sharing the purely electronic mechanism of pairing~\cite{Sharma2020} provided by the interaction with acoustic plasmons~\cite{Frohlich1968,Canright1989,Ruhman2017,Fatemi2018}. In this case the absence of isotope effect will serve as a hallmark of it~\cite{Ruvalds1981}. The emergence of acoustic plasmons in WSMs introduce new mechanisms of plasmon instability in materials with linear quasiparticle dispersion. Due to the lack of Galilean invariance and singularities in ${\rm Re}\,\varepsilon(\omega,{\bf q})$ at quasiparticle dispersion line $\omega=vq$, negative Landau damping in one-component Dirac plasma is prohibited~\cite{Svintsov2019_PRL} in collisionless regime and the emission of plasmons is feasible only when the hydrodynamic transport take place~\cite{Svintsov2019_PRB}. In heterogeneous two-component Weyl/Dirac plasma the possibility of ${\rm \check{C}}$herenkov radiation of acoustic plasmons~\cite{PinesSchrieffer1962} is determined only by the relation between the drift velocity of the fast component and the "sound" velocity $s$, being independent on the particular transport regime. Additionally, the emergence of acoustic plasmons could qualitatively change the renormalization of the quasiparticle band structure due to the strong plasmon-WF coupling. In addition to predicted plasmaronic subband~\cite{Hofmann2015_Liquid} associated with optical plasmons, the composite quasiparticles consisting of acoustic plasmon and WF - Weyl soundarons~\cite{Principi2011} - may arise and mix with each other. 

\section*{Acknowledgments}
This work was supported by Grant 16-19-10557 of Russian Science Foundation. A. N. Afanasiev acknowledges the support by the Foundation for the Advancement of Theoretical Physics and Mathematics “BASIS” (grant 19-1-5-127-1).

\appendix

\section{Explicit solution for the self-consistent response to the initial perturbation in isotropic degenerate two-component WP}
\label{Sec_AP:Rel}

The time evolution of the self-consistent response of the form
\begin{equation}
    n_j({\bf r},{\bf p},t)=n_j^{(0)}(p)+\delta n_j({\bf r},{\bf p},t)
\end{equation}
to the weak initial perturbation
\begin{equation}
    g({\bf r},{\bf p})=n_2(0,{\bf r}, {\bf p})-n_2^{(0)}(p)
\end{equation}
is governed by the set of linearized collisionless Boltzman equations for both components 
\begin{equation}
    \label{Kin_eq}
    \frac{\partial \delta n_j}{\partial t}+{\bf v}_j \frac{\partial \delta n_j}{\partial {\bf r}}+{\bf \nabla}\mathcal{V}\frac{\partial n_j^{(0)}}{\partial {\bf p}}=0
\end{equation}
where $j=1,2$ is the plasma component index, $n_j^{(0)}(p)=\theta(p_{F}^{(j)}-p)$ is the equilibrium distribution function, $\delta n_j ({\bf r},{\bf p},t) \ll n_j^{(0)}(p)$ is the small non-equilibrium correction to it, ${\bf p}=\hbar {\bf k}$ is the momentum of WFs. In this section we re-introduce the Planck constant for the sake of clarity. According to~(\ref{Kin_eq}) the spatial Fourier transform of the self-consistent potential $\mathcal{V}({\bf r},t)$ is given by
\begin{gather}
    \label{Eq_Ap:RelDynamicsCommon}
    \mathcal{V}_{q}(t)=\frac{1}{2\pi}\int\limits_{-\infty}^{+\infty}{\rm e}^{-i\omega t}\mathcal{V}^{(+)}_{\omega {q}} d\omega\\
    \frac{\mathcal{V}^{(+)}_{\omega{q}}}{V_q}=-\frac{\eta_2}{i\,\varepsilon(\omega,{q})}\int\frac{g_{q}({\bf p})}{{\bf v}_2{\bf q}-\omega-i 0}\frac{d{\bf p}}{(2\pi\hbar)^3},
    \label{Eq_Ap:RelSpectraCommon}
\end{gather}
where $V_q=\frac{4\pi e^2}{\varkappa_b q^2}$ is the three-dimensional Coulomb potential. We consider the initial perturbation to be plane wave-like $g({\bf r},{\bf p})=g_{q}({\bf p}) {\rm e}^{i{\bf q r}}$. Since equation~(\ref{Kin_eq}) describes the kinetics of degenerate plasma in quasiclassical limit $v_2 p \ll \mu_2$, $\hbar \omega \ll \mu_2$, the dielectric function in~(\ref{Eq_Ap:RelSpectraCommon}) determining the self-consistent response coincide with the one given by Eqs.~(\ref{Eq:TwoCompEpsilon}) and~(\ref{Eq:QuasiclassicalEpsilon}). For the sake of simplicity we consider the perturbation to be isotropic in momentum space. Therefore, integrations over magnitude $p=|{\bf p}|$ and directions ${\bf e_p}$ of the WF's momenta can be carried out separately for an arbitrary form of $g_{q}(p)$. Combining~(\ref{Eq_Ap:RelDynamicsCommon}) and~(\ref{Eq_Ap:RelSpectraCommon}) at $g_{q}({\bf p})=g_{q}(p)$, the time evolution of the self-consistent potential in WP per one of the  $\mathcal{N}_2({q})=\eta_2\int g_{q}(p) \frac{d{\bf p}}{(2\pi\hbar)^3}$ perturbed WFs is given by
\begin{equation}
    \delta\mathcal{V}_{q}(t)=\frac{\mathcal{V}_{q}(t)}{V_q\mathcal{N}_2({q})}=\int\limits_{-\infty}^{+\infty}\frac{d\omega}{2\pi}{\rm e}^{-i \omega t} \tilde{K}_2(\omega,q)
\end{equation}
is determined by the quasiclassical form of the dimensionless PRA relaxation propagator
\begin{equation}
    \tilde{K}_2(\omega,q)=\frac{1}{\varepsilon(\omega,q)}\frac{\tilde{\Pi}_2(0,q)-\tilde{\Pi}_2(\omega,q)}{i\omega},
\end{equation}
where the dimensionless non-interacting quasiclassical polarizability $\tilde{\Pi}_2(\omega,q)=\Pi_2(\omega,q)/D_2(\mu_2)$ has the form
\begin{multline}
    \label{Eq_Ap:NonIntPolarizability}
    \tilde{\Pi}_2(\omega,q)=-\left[1+\frac{\omega}{2 v_2 q}\ln\left|\frac{\omega-v_2 q}{\omega+v_2 q}\right|\right]-\\
    -i \pi \frac{\omega}{2 v_2 q} \theta(v_2 q - \omega)\theta(v_2 q + \omega).
\end{multline}
Here $D_2(\mu_2)$ is the density of states of the fast WF at Fermi level.


\section{Effect of the internode chirality-flip transitions on the dielectric function of WSM}
\label{Sec_Ap:Chirality}

Dielectric function of the multinode WSM with the allowance for the chirality-flip transitions between the nearest Weyl nodes of the same type has the form
\begin{equation}
    \Delta\varepsilon(\omega,\textbf{q})=\sum\limits_{\substack{\chi,\chi' \\ n,i}}\Delta\epsilon_{n,i}^{(\chi\chi')}(\omega,\textbf{q})
\end{equation}
where $\chi=\pm$, $\chi'=\pm$ are chiralities, $i$ denote the symmetry operations in WSM and $n$ is the node group index. Non-diagonal (in $\chi$) components of $\Delta\epsilon_{n,i}^{(\chi\chi')}(\omega,\textbf{q})$ correspond to chirality-flip transitions, while diagonal ones coincide with the terms of~(\ref{Eq:EpsilonWSM})
\begin{equation}
    \Delta\epsilon_{n,i}^{(\chi\chi)}(\omega,\textbf{q})=\Delta\varepsilon_{n}(\omega,\left|v_n^{-1}\hat{v}_{n}g_{i}{\bf q}\right|),
\end{equation}
where we have already neglected the tilt velocity induced Doppler shift for the same reasons stated in Sec.~\ref{Sec:PlSpec}.

Chiral-dependent Hamiltonian relevant to a certain node $W^{(\chi)}_{n,i}$ is derived from the Hamiltonian of the isolated Weyl node $H=\hat{v}_{n,i}\textbf{k} \cdot \boldsymbol{\sigma}$ by substitutions $\textbf{k}\rightarrow \textbf{k}+{
\bf K}_{W_{n,i}^{(\chi)}}$ and $\hat{\sigma}\rightarrow \chi\hat{\sigma}$, where $\hat{v}_{n,i}=\hat{v}_n g_i$ is the local Weyl velocity tensor and ${\bf K}_{W_{n,i}^{(\chi)}}$ is the position of the Weyl node in the Brillouin zone. The neighbor Weyl nodes   with opposite chiralities are separated by the ${\bf K}_{W_{n,i}^{(+)}}-{\bf K}_{W_{n,i}^{(-)}}={\bf Q}_{n,i}=g_{i}{\bf Q}_n$ specific for the group $W_n$. Therefore the components of chirality-dependent dielectric function are connected to diagonal ones via the transformation
\begin{equation}
    \Delta\epsilon_{n,i}^{(\chi\chi')}(\omega,{\bf q})=\Delta\epsilon_{n,i}^{(\chi\chi)}\left(\omega,{\bf q}+\frac{\chi-\chi'}{2}{\bf Q}_{n,i}\right)
\end{equation}
leading to
\begin{equation}
    \Delta\epsilon^{(-\chi\chi)}_{n,i}(\omega,\textbf{q})=\Delta\varepsilon^{(\chi\chi)}_{n,i}(\omega,\textbf{q}+\chi\textbf{Q}_{n,i}),
\end{equation}
and the equivalence of diagonal components $\Delta\epsilon_{n,i}^{(++)}(\omega,{\bf q})=\Delta\epsilon_{n,i}^{(--)}(\omega,{\bf q})$ (in equilibrium). 

Finally, the total dielectric function is given by the sum of intranode term coinciding with~(\ref{Eq:EpsilonWSM}) and the internode one
\begin{gather}
    \varepsilon(\omega,\textbf{q})=\varepsilon_{\rm intra}(\omega,\textbf{q})+\varepsilon_{\rm inter}(\omega,\textbf{q})\\
    \varepsilon_{\rm intra}(\omega,\textbf{q})=1+\sum\limits_{\substack{\chi=\pm \\ n,i}}\Delta\epsilon_{n,i}^{(\chi\chi)}(\omega,\textbf{q})\\
    \varepsilon_{\rm inter}(\omega,\textbf{q})=\sum\limits_{\substack{\chi=\pm \\ n,i}}\Delta\epsilon_{n,i}^{(\chi\chi)}(\omega,\textbf{q}+\chi\textbf{Q}_{n,i}).
\end{gather}
Since the plasmon dispersion in the short wavelength domain is limited by the critical endpoints (see Sec.~\ref{Sec:ShortWL}) and thus we are interested only in $\omega/v_{n,i}({\bf e_q}),|{\bf q}|\leq k_F^{(n,i)}({\bf e_q})$, the resolvability condition $k_F^{(n,i)}({\bf e}_{_{{\bf Q}_{n,i}}})\ll |{\bf Q}_{n,i}|$ of the Weyl nodes leads to 
\begin{equation}
\Delta\epsilon^{(\chi\chi)}_{n,i}(\omega,\textbf{q}+\chi\textbf{Q}_{n,i})\approx \Delta\epsilon^{(\chi\chi)}_{n,i}(0,|\textbf{Q}_{n,i}|),
\end{equation}
which is given predominantly by the interband transitions (typical intraband excitation energy $\mu_{n}$ is incomparable with $v({\bf e}_{_{{\bf Q}_{n,i}}}) |{\bf Q}_{n,i}|$). Therefore, the effect of internodal scattering reduces to additional quite small renormalization of background dielectric constant 
\begin{equation}
    \varkappa_b\varkappa_0\rightarrow  \varkappa_b\varkappa_0 \left[ 1+\frac{\eta_1\alpha_1}{3\pi}\ln\frac{\Lambda}{Q_1}+\frac{\eta_2\alpha_2}{3\pi}\ln\frac{\Lambda}{Q_2}\right]
\end{equation}
where $\Lambda$ is the cutoff wave vector of linear dispersion.

\bibliography{Bib_WSM_Corr}

\begin{thebibliography}{56}%
\makeatletter
\providecommand \@ifxundefined [1]{%
 \@ifx{#1\undefined}
}%
\providecommand \@ifnum [1]{%
 \ifnum #1\expandafter \@firstoftwo
 \else \expandafter \@secondoftwo
 \fi
}%
\providecommand \@ifx [1]{%
 \ifx #1\expandafter \@firstoftwo
 \else \expandafter \@secondoftwo
 \fi
}%
\providecommand \natexlab [1]{#1}%
\providecommand \enquote  [1]{``#1''}%
\providecommand \bibnamefont  [1]{#1}%
\providecommand \bibfnamefont [1]{#1}%
\providecommand \citenamefont [1]{#1}%
\providecommand \href@noop [0]{\@secondoftwo}%
\providecommand \href [0]{\begingroup \@sanitize@url \@href}%
\providecommand \@href[1]{\@@startlink{#1}\@@href}%
\providecommand \@@href[1]{\endgroup#1\@@endlink}%
\providecommand \@sanitize@url [0]{\catcode `\\12\catcode `\$12\catcode
  `\&12\catcode `\#12\catcode `\^12\catcode `\_12\catcode `\%12\relax}%
\providecommand \@@startlink[1]{}%
\providecommand \@@endlink[0]{}%
\providecommand \url  [0]{\begingroup\@sanitize@url \@url }%
\providecommand \@url [1]{\endgroup\@href {#1}{\urlprefix }}%
\providecommand \urlprefix  [0]{URL }%
\providecommand \Eprint [0]{\href }%
\providecommand \doibase [0]{https://doi.org/}%
\providecommand \selectlanguage [0]{\@gobble}%
\providecommand \bibinfo  [0]{\@secondoftwo}%
\providecommand \bibfield  [0]{\@secondoftwo}%
\providecommand \translation [1]{[#1]}%
\providecommand \BibitemOpen [0]{}%
\providecommand \bibitemStop [0]{}%
\providecommand \bibitemNoStop [0]{.\EOS\space}%
\providecommand \EOS [0]{\spacefactor3000\relax}%
\providecommand \BibitemShut  [1]{\csname bibitem#1\endcsname}%
\let\auto@bib@innerbib\@empty
\bibitem [{\citenamefont {Weyl}(1929)}]{Weyl1929}%
  \BibitemOpen
  \bibfield  {author} {\bibinfo {author} {\bibfnamefont {H.}~\bibnamefont
  {Weyl}},\ }\bibfield  {title} {\bibinfo {title} {Gravitation and the
  electron},\ }\href {https://doi.org/10.1073/pnas.15.4.323} {\bibfield
  {journal} {\bibinfo  {journal} {Proc. Natl. Acad. Sci. U.S.A.}\ }\textbf
  {\bibinfo {volume} {15}},\ \bibinfo {pages} {323} (\bibinfo {year}
  {1929})}\BibitemShut {NoStop}%
\bibitem [{\citenamefont {{A}rmitage}\ \emph {et~al.}(2018)\citenamefont
  {{A}rmitage}, \citenamefont {Mele},\ and\ \citenamefont
  {Vishwanath}}]{Armitage2018}%
  \BibitemOpen
  \bibfield  {author} {\bibinfo {author} {\bibfnamefont {N.~P.}\ \bibnamefont
  {{A}rmitage}}, \bibinfo {author} {\bibfnamefont {E.~J.}\ \bibnamefont
  {Mele}},\ and\ \bibinfo {author} {\bibfnamefont {A.}~\bibnamefont
  {Vishwanath}},\ }\bibfield  {title} {\bibinfo {title} {{W}eyl and {D}irac
  semimetals in three-dimensional solids},\ }\href
  {https://doi.org/10.1103/RevModPhys.90.015001} {\bibfield  {journal}
  {\bibinfo  {journal} {Rev. Mod. Phys.}\ }\textbf {\bibinfo {volume} {90}},\
  \bibinfo {pages} {015001} (\bibinfo {year} {2018})}\BibitemShut {NoStop}%
\bibitem [{\citenamefont {Burkov}(2018)}]{Burkov2018}%
  \BibitemOpen
  \bibfield  {author} {\bibinfo {author} {\bibfnamefont {A.}~\bibnamefont
  {Burkov}},\ }\bibfield  {title} {\bibinfo {title} {Weyl {M}etals},\ }\href
  {https://doi.org/10.1146/annurev-conmatphys-033117-054129} {\bibfield
  {journal} {\bibinfo  {journal} {Annual Review of Condensed Matter Physics}\
  }\textbf {\bibinfo {volume} {9}},\ \bibinfo {pages} {359} (\bibinfo {year}
  {2018})},\ \Eprint
  {https://arxiv.org/abs/https://doi.org/10.1146/annurev-conmatphys-033117-054129}
  {https://doi.org/10.1146/annurev-conmatphys-033117-054129} \BibitemShut
  {NoStop}%
\bibitem [{\citenamefont {Das~Sarma}\ and\ \citenamefont
  {Hwang}(2009)}]{DasSarma2009}%
  \BibitemOpen
  \bibfield  {author} {\bibinfo {author} {\bibfnamefont {S.}~\bibnamefont
  {Das~Sarma}}\ and\ \bibinfo {author} {\bibfnamefont {E.~H.}\ \bibnamefont
  {Hwang}},\ }\bibfield  {title} {\bibinfo {title} {Collective {M}odes of the
  {M}assless {D}irac {P}lasma},\ }\href
  {https://doi.org/10.1103/PhysRevLett.102.206412} {\bibfield  {journal}
  {\bibinfo  {journal} {Phys. Rev. Lett.}\ }\textbf {\bibinfo {volume} {102}},\
  \bibinfo {pages} {206412} (\bibinfo {year} {2009})}\BibitemShut {NoStop}%
\bibitem [{\citenamefont {Sachdeva}\ \emph {et~al.}(2015)\citenamefont
  {Sachdeva}, \citenamefont {Thakur}, \citenamefont {Vignale},\ and\
  \citenamefont {Agarwal}}]{Sachdeva2015}%
  \BibitemOpen
  \bibfield  {author} {\bibinfo {author} {\bibfnamefont {R.}~\bibnamefont
  {Sachdeva}}, \bibinfo {author} {\bibfnamefont {A.}~\bibnamefont {Thakur}},
  \bibinfo {author} {\bibfnamefont {G.}~\bibnamefont {Vignale}},\ and\ \bibinfo
  {author} {\bibfnamefont {A.}~\bibnamefont {Agarwal}},\ }\bibfield  {title}
  {\bibinfo {title} {Plasmon modes of a massive {D}irac plasma, and their
  superlattices},\ }\href {https://doi.org/10.1103/PhysRevB.91.205426}
  {\bibfield  {journal} {\bibinfo  {journal} {Phys. Rev. B}\ }\textbf {\bibinfo
  {volume} {91}},\ \bibinfo {pages} {205426} (\bibinfo {year}
  {2015})}\BibitemShut {NoStop}%
\bibitem [{\citenamefont {Silin}(1960)}]{Silin1960}%
  \BibitemOpen
  \bibfield  {author} {\bibinfo {author} {\bibfnamefont {V.~P.}\ \bibnamefont
  {Silin}},\ }\bibfield  {title} {\bibinfo {title} {On the {E}lectromagnetic
  {P}roperties of a {R}elativistic {P}lasma},\ }\href
  {http://www.jetp.ac.ru/cgi-bin/e/index/e/11/5/p1136?a=list} {\bibfield
  {journal} {\bibinfo  {journal} {Soviet Phys. JETP}\ }\textbf {\bibinfo
  {volume} {11}},\ \bibinfo {pages} {1136} (\bibinfo {year}
  {1960})}\BibitemShut {NoStop}%
\bibitem [{\citenamefont {Lv}\ and\ \citenamefont
  {Zhang}(2013)}]{Lv2013_Theory}%
  \BibitemOpen
  \bibfield  {author} {\bibinfo {author} {\bibfnamefont {M.}~\bibnamefont
  {Lv}}\ and\ \bibinfo {author} {\bibfnamefont {S.-C.}\ \bibnamefont {Zhang}},\
  }\bibfield  {title} {\bibinfo {title} {Dielectric function, {F}riedel
  oscillation and plasmons in {W}eyl semimetals},\ }\href
  {https://doi.org/10.1142/S0217979213501774} {\bibfield  {journal} {\bibinfo
  {journal} {Int. J. Mod. Phys. B}\ }\textbf {\bibinfo {volume} {27}},\
  \bibinfo {pages} {1350177} (\bibinfo {year} {2013})}\BibitemShut {NoStop}%
\bibitem [{\citenamefont {Hofmann}\ and\ \citenamefont
  {Das~Sarma}(2015)}]{Hofmann2015_Signature}%
  \BibitemOpen
  \bibfield  {author} {\bibinfo {author} {\bibfnamefont {J.}~\bibnamefont
  {Hofmann}}\ and\ \bibinfo {author} {\bibfnamefont {S.}~\bibnamefont
  {Das~Sarma}},\ }\bibfield  {title} {\bibinfo {title} {Plasmon signature in
  {D}irac-{W}eyl liquids},\ }\href {https://doi.org/10.1103/PhysRevB.91.241108}
  {\bibfield  {journal} {\bibinfo  {journal} {Phys. Rev. B}\ }\textbf {\bibinfo
  {volume} {91}},\ \bibinfo {pages} {241108} (\bibinfo {year}
  {2015})}\BibitemShut {NoStop}%
\bibitem [{\citenamefont {Thakur}\ \emph {et~al.}(2017)\citenamefont {Thakur},
  \citenamefont {Sachdeva},\ and\ \citenamefont {Agarwal}}]{Thakur2017}%
  \BibitemOpen
  \bibfield  {author} {\bibinfo {author} {\bibfnamefont {A.}~\bibnamefont
  {Thakur}}, \bibinfo {author} {\bibfnamefont {R.}~\bibnamefont {Sachdeva}},\
  and\ \bibinfo {author} {\bibfnamefont {A.}~\bibnamefont {Agarwal}},\
  }\bibfield  {title} {\bibinfo {title} {Dynamical polarizability, screening
  and plasmons in one, two and three dimensional massive {D}irac systems},\
  }\href {https://doi.org/https://doi.org/10.1088/1361-648X/aa57bd} {\bibfield
  {journal} {\bibinfo  {journal} {J. Phys.: Condens. Matter}\ }\textbf
  {\bibinfo {volume} {29}},\ \bibinfo {pages} {105701} (\bibinfo {year}
  {2017})}\BibitemShut {NoStop}%
\bibitem [{\citenamefont {Zhou}\ \emph {et~al.}(2015)\citenamefont {Zhou},
  \citenamefont {Chang},\ and\ \citenamefont {Xiao}}]{Zhou2015}%
  \BibitemOpen
  \bibfield  {author} {\bibinfo {author} {\bibfnamefont {J.}~\bibnamefont
  {Zhou}}, \bibinfo {author} {\bibfnamefont {H.-R.}\ \bibnamefont {Chang}},\
  and\ \bibinfo {author} {\bibfnamefont {D.}~\bibnamefont {Xiao}},\ }\bibfield
  {title} {\bibinfo {title} {Plasmon mode as a detection of the chiral anomaly
  in {W}eyl semimetals},\ }\href {https://doi.org/10.1103/PhysRevB.91.035114}
  {\bibfield  {journal} {\bibinfo  {journal} {Phys. Rev. B}\ }\textbf {\bibinfo
  {volume} {91}},\ \bibinfo {pages} {035114} (\bibinfo {year}
  {2015})}\BibitemShut {NoStop}%
\bibitem [{\citenamefont {Song}\ and\ \citenamefont {Rudner}(2017)}]{Song2017}%
  \BibitemOpen
  \bibfield  {author} {\bibinfo {author} {\bibfnamefont {J.~C.~W.}\
  \bibnamefont {Song}}\ and\ \bibinfo {author} {\bibfnamefont {M.~S.}\
  \bibnamefont {Rudner}},\ }\bibfield  {title} {\bibinfo {title} {Fermi arc
  plasmons in {W}eyl semimetals},\ }\href
  {https://doi.org/10.1103/PhysRevB.96.205443} {\bibfield  {journal} {\bibinfo
  {journal} {Phys. Rev. B}\ }\textbf {\bibinfo {volume} {96}},\ \bibinfo
  {pages} {205443} (\bibinfo {year} {2017})}\BibitemShut {NoStop}%
\bibitem [{\citenamefont {Andolina}\ \emph {et~al.}(2018)\citenamefont
  {Andolina}, \citenamefont {Pellegrino}, \citenamefont {Koppens},\ and\
  \citenamefont {Polini}}]{Andolina2018}%
  \BibitemOpen
  \bibfield  {author} {\bibinfo {author} {\bibfnamefont {G.~M.}\ \bibnamefont
  {Andolina}}, \bibinfo {author} {\bibfnamefont {F.~M.~D.}\ \bibnamefont
  {Pellegrino}}, \bibinfo {author} {\bibfnamefont {F.~H.~L.}\ \bibnamefont
  {Koppens}},\ and\ \bibinfo {author} {\bibfnamefont {M.}~\bibnamefont
  {Polini}},\ }\bibfield  {title} {\bibinfo {title} {Quantum nonlocal theory of
  topological {F}ermi arc plasmons in {W}eyl semimetals},\ }\href
  {https://doi.org/10.1103/PhysRevB.97.125431} {\bibfield  {journal} {\bibinfo
  {journal} {Phys. Rev. B}\ }\textbf {\bibinfo {volume} {97}},\ \bibinfo
  {pages} {125431} (\bibinfo {year} {2018})}\BibitemShut {NoStop}%
\bibitem [{\citenamefont {Adinehvand}\ \emph {et~al.}(2019)\citenamefont
  {Adinehvand}, \citenamefont {Faraei}, \citenamefont {Farajollahpour},\ and\
  \citenamefont {Jafari}}]{Adinehvand2019}%
  \BibitemOpen
  \bibfield  {author} {\bibinfo {author} {\bibfnamefont {F.}~\bibnamefont
  {Adinehvand}}, \bibinfo {author} {\bibfnamefont {Z.}~\bibnamefont {Faraei}},
  \bibinfo {author} {\bibfnamefont {T.}~\bibnamefont {Farajollahpour}},\ and\
  \bibinfo {author} {\bibfnamefont {S.~A.}\ \bibnamefont {Jafari}},\ }\bibfield
   {title} {\bibinfo {title} {Sound of {F}ermi arcs: a linearly dispersing
  gapless surface plasmon mode in undoped {W}eyl semimetals},\ }\href
  {https://doi.org/10.1103/PhysRevB.100.195408} {\bibfield  {journal} {\bibinfo
   {journal} {Phys. Rev. B}\ }\textbf {\bibinfo {volume} {100}},\ \bibinfo
  {pages} {195408} (\bibinfo {year} {2019})}\BibitemShut {NoStop}%
\bibitem [{\citenamefont {Hofmann}\ and\ \citenamefont
  {Das~Sarma}(2016)}]{Hofmann2016}%
  \BibitemOpen
  \bibfield  {author} {\bibinfo {author} {\bibfnamefont {J.}~\bibnamefont
  {Hofmann}}\ and\ \bibinfo {author} {\bibfnamefont {S.}~\bibnamefont
  {Das~Sarma}},\ }\bibfield  {title} {\bibinfo {title} {Surface plasmon
  polaritons in topological {W}eyl semimetals},\ }\href
  {https://doi.org/10.1103/PhysRevB.93.241402} {\bibfield  {journal} {\bibinfo
  {journal} {Phys. Rev. B}\ }\textbf {\bibinfo {volume} {93}},\ \bibinfo
  {pages} {241402} (\bibinfo {year} {2016})}\BibitemShut {NoStop}%
\bibitem [{\citenamefont {Chiarello}\ \emph {et~al.}(2019)\citenamefont
  {Chiarello}, \citenamefont {Hofmann}, \citenamefont {Li}, \citenamefont
  {Fabio}, \citenamefont {Guo}, \citenamefont {Chen}, \citenamefont
  {Das~Sarma},\ and\ \citenamefont {Politano}}]{Chiarello2019}%
  \BibitemOpen
  \bibfield  {author} {\bibinfo {author} {\bibfnamefont {G.}~\bibnamefont
  {Chiarello}}, \bibinfo {author} {\bibfnamefont {J.}~\bibnamefont {Hofmann}},
  \bibinfo {author} {\bibfnamefont {Z.}~\bibnamefont {Li}}, \bibinfo {author}
  {\bibfnamefont {V.}~\bibnamefont {Fabio}}, \bibinfo {author} {\bibfnamefont
  {L.}~\bibnamefont {Guo}}, \bibinfo {author} {\bibfnamefont {X.}~\bibnamefont
  {Chen}}, \bibinfo {author} {\bibfnamefont {S.}~\bibnamefont {Das~Sarma}},\
  and\ \bibinfo {author} {\bibfnamefont {A.}~\bibnamefont {Politano}},\
  }\bibfield  {title} {\bibinfo {title} {Tunable surface plasmons in {W}eyl
  semimetals {T}a{A}s and {N}b{A}s},\ }\href
  {https://doi.org/10.1103/PhysRevB.99.121401} {\bibfield  {journal} {\bibinfo
  {journal} {Phys. Rev. B}\ }\textbf {\bibinfo {volume} {99}},\ \bibinfo
  {pages} {121401} (\bibinfo {year} {2019})}\BibitemShut {NoStop}%
\bibitem [{\citenamefont {Lupi}\ and\ \citenamefont {Molle}(2020)}]{Lupi2020}%
  \BibitemOpen
  \bibfield  {author} {\bibinfo {author} {\bibfnamefont {S.}~\bibnamefont
  {Lupi}}\ and\ \bibinfo {author} {\bibfnamefont {A.}~\bibnamefont {Molle}},\
  }\bibfield  {title} {\bibinfo {title} {Emerging {D}irac materials for {TH}z
  plasmonics},\ }\href
  {https://doi.org/https://doi.org/10.1016/j.apmt.2020.100732} {\bibfield
  {journal} {\bibinfo  {journal} {Applied Materials Today}\ }\textbf {\bibinfo
  {volume} {20}},\ \bibinfo {pages} {100732} (\bibinfo {year}
  {2020})}\BibitemShut {NoStop}%
\bibitem [{\citenamefont {Lv}\ \emph {et~al.}(2015)\citenamefont {Lv},
  \citenamefont {{W}eng}, \citenamefont {Fu}, \citenamefont {{W}ang},
  \citenamefont {Miao}, \citenamefont {Ma}, \citenamefont {Richard},
  \citenamefont {Huang}, \citenamefont {Zhao}, \citenamefont {Chen},
  \citenamefont {Fang}, \citenamefont {{D}ai}, \citenamefont {Qian},\ and\
  \citenamefont {{D}ing}}]{Lv2015TaAs}%
  \BibitemOpen
  \bibfield  {author} {\bibinfo {author} {\bibfnamefont {B.~Q.}\ \bibnamefont
  {Lv}}, \bibinfo {author} {\bibfnamefont {H.~M.}\ \bibnamefont {{W}eng}},
  \bibinfo {author} {\bibfnamefont {B.~B.}\ \bibnamefont {Fu}}, \bibinfo
  {author} {\bibfnamefont {X.~P.}\ \bibnamefont {{W}ang}}, \bibinfo {author}
  {\bibfnamefont {H.}~\bibnamefont {Miao}}, \bibinfo {author} {\bibfnamefont
  {J.}~\bibnamefont {Ma}}, \bibinfo {author} {\bibfnamefont {P.}~\bibnamefont
  {Richard}}, \bibinfo {author} {\bibfnamefont {X.~C.}\ \bibnamefont {Huang}},
  \bibinfo {author} {\bibfnamefont {L.~X.}\ \bibnamefont {Zhao}}, \bibinfo
  {author} {\bibfnamefont {G.~F.}\ \bibnamefont {Chen}}, \bibinfo {author}
  {\bibfnamefont {Z.}~\bibnamefont {Fang}}, \bibinfo {author} {\bibfnamefont
  {X.}~\bibnamefont {{D}ai}}, \bibinfo {author} {\bibfnamefont
  {T.}~\bibnamefont {Qian}},\ and\ \bibinfo {author} {\bibfnamefont
  {H.}~\bibnamefont {{D}ing}},\ }\bibfield  {title} {\bibinfo {title}
  {Experimental discovery of {W}eyl semimetal {T}a{A}s},\ }\href
  {https://doi.org/10.1103/PhysRevX.5.031013} {\bibfield  {journal} {\bibinfo
  {journal} {Phys. Rev. X}\ }\textbf {\bibinfo {volume} {5}},\ \bibinfo {pages}
  {031013} (\bibinfo {year} {2015})}\BibitemShut {NoStop}%
\bibitem [{\citenamefont {Xu}\ \emph {et~al.}(2015{\natexlab{a}})\citenamefont
  {Xu}, \citenamefont {Belopolski}, \citenamefont {Sanchez}, \citenamefont
  {Zhang}, \citenamefont {Chang}, \citenamefont {Guo}, \citenamefont {Bian},
  \citenamefont {Yuan}, \citenamefont {Lu}, \citenamefont {Chang},
  \citenamefont {Shibayev}, \citenamefont {Prokopovych}, \citenamefont
  {{A}lidoust}, \citenamefont {Zheng}, \citenamefont {Lee}, \citenamefont
  {Huang}, \citenamefont {Sankar}, \citenamefont {Chou}, \citenamefont {Hsu},
  \citenamefont {Jeng}, \citenamefont {Bansil}, \citenamefont {Neupert},
  \citenamefont {Strocov}, \citenamefont {Lin}, \citenamefont {Jia},\ and\
  \citenamefont {Hasan}}]{Xu2015TaP}%
  \BibitemOpen
  \bibfield  {author} {\bibinfo {author} {\bibfnamefont {S.-Y.}\ \bibnamefont
  {Xu}}, \bibinfo {author} {\bibfnamefont {I.}~\bibnamefont {Belopolski}},
  \bibinfo {author} {\bibfnamefont {D.~S.}\ \bibnamefont {Sanchez}}, \bibinfo
  {author} {\bibfnamefont {C.}~\bibnamefont {Zhang}}, \bibinfo {author}
  {\bibfnamefont {G.}~\bibnamefont {Chang}}, \bibinfo {author} {\bibfnamefont
  {C.}~\bibnamefont {Guo}}, \bibinfo {author} {\bibfnamefont {G.}~\bibnamefont
  {Bian}}, \bibinfo {author} {\bibfnamefont {Z.}~\bibnamefont {Yuan}}, \bibinfo
  {author} {\bibfnamefont {H.}~\bibnamefont {Lu}}, \bibinfo {author}
  {\bibfnamefont {T.-R.}\ \bibnamefont {Chang}}, \bibinfo {author}
  {\bibfnamefont {P.~P.}\ \bibnamefont {Shibayev}}, \bibinfo {author}
  {\bibfnamefont {M.~L.}\ \bibnamefont {Prokopovych}}, \bibinfo {author}
  {\bibfnamefont {N.}~\bibnamefont {{A}lidoust}}, \bibinfo {author}
  {\bibfnamefont {H.}~\bibnamefont {Zheng}}, \bibinfo {author} {\bibfnamefont
  {C.-C.}\ \bibnamefont {Lee}}, \bibinfo {author} {\bibfnamefont {S.-M.}\
  \bibnamefont {Huang}}, \bibinfo {author} {\bibfnamefont {R.}~\bibnamefont
  {Sankar}}, \bibinfo {author} {\bibfnamefont {F.}~\bibnamefont {Chou}},
  \bibinfo {author} {\bibfnamefont {C.-H.}\ \bibnamefont {Hsu}}, \bibinfo
  {author} {\bibfnamefont {H.-T.}\ \bibnamefont {Jeng}}, \bibinfo {author}
  {\bibfnamefont {A.}~\bibnamefont {Bansil}}, \bibinfo {author} {\bibfnamefont
  {T.}~\bibnamefont {Neupert}}, \bibinfo {author} {\bibfnamefont {V.~N.}\
  \bibnamefont {Strocov}}, \bibinfo {author} {\bibfnamefont {H.}~\bibnamefont
  {Lin}}, \bibinfo {author} {\bibfnamefont {S.}~\bibnamefont {Jia}},\ and\
  \bibinfo {author} {\bibfnamefont {M.~Z.}\ \bibnamefont {Hasan}},\ }\bibfield
  {title} {\bibinfo {title} {Experimental discovery of a topological {W}eyl
  semimetal state in {T}a{P}},\ }\href {https://doi.org/10.1126/sciadv.1501092}
  {\bibfield  {journal} {\bibinfo  {journal} {Sci. {A}dv.}\ }\textbf {\bibinfo
  {volume} {1}},\ \bibinfo {pages} {1501092} (\bibinfo {year}
  {2015}{\natexlab{a}})}\BibitemShut {NoStop}%
\bibitem [{\citenamefont {Xu}\ \emph {et~al.}(2015{\natexlab{b}})\citenamefont
  {Xu}, \citenamefont {Belopolski}, \citenamefont {{A}lidoust}, \citenamefont
  {Neupane}, \citenamefont {Bian}, \citenamefont {Zhang}, \citenamefont
  {Sankar}, \citenamefont {Chang}, \citenamefont {Yuan}, \citenamefont {Lee},
  \citenamefont {Huang}, \citenamefont {Zheng}, \citenamefont {Ma},
  \citenamefont {Sanchez}, \citenamefont {{W}ang}, \citenamefont {Bansil},
  \citenamefont {Chou}, \citenamefont {Shibayev}, \citenamefont {Lin},
  \citenamefont {Jia},\ and\ \citenamefont {Hasan}}]{Xu2015TaAs}%
  \BibitemOpen
  \bibfield  {author} {\bibinfo {author} {\bibfnamefont {S.-Y.}\ \bibnamefont
  {Xu}}, \bibinfo {author} {\bibfnamefont {I.}~\bibnamefont {Belopolski}},
  \bibinfo {author} {\bibfnamefont {N.}~\bibnamefont {{A}lidoust}}, \bibinfo
  {author} {\bibfnamefont {M.}~\bibnamefont {Neupane}}, \bibinfo {author}
  {\bibfnamefont {G.}~\bibnamefont {Bian}}, \bibinfo {author} {\bibfnamefont
  {C.}~\bibnamefont {Zhang}}, \bibinfo {author} {\bibfnamefont
  {R.}~\bibnamefont {Sankar}}, \bibinfo {author} {\bibfnamefont
  {G.}~\bibnamefont {Chang}}, \bibinfo {author} {\bibfnamefont
  {Z.}~\bibnamefont {Yuan}}, \bibinfo {author} {\bibfnamefont {C.-C.}\
  \bibnamefont {Lee}}, \bibinfo {author} {\bibfnamefont {S.-M.}\ \bibnamefont
  {Huang}}, \bibinfo {author} {\bibfnamefont {H.}~\bibnamefont {Zheng}},
  \bibinfo {author} {\bibfnamefont {J.}~\bibnamefont {Ma}}, \bibinfo {author}
  {\bibfnamefont {D.~S.}\ \bibnamefont {Sanchez}}, \bibinfo {author}
  {\bibfnamefont {B.}~\bibnamefont {{W}ang}}, \bibinfo {author} {\bibfnamefont
  {A.}~\bibnamefont {Bansil}}, \bibinfo {author} {\bibfnamefont
  {F.}~\bibnamefont {Chou}}, \bibinfo {author} {\bibfnamefont {P.~P.}\
  \bibnamefont {Shibayev}}, \bibinfo {author} {\bibfnamefont {H.}~\bibnamefont
  {Lin}}, \bibinfo {author} {\bibfnamefont {S.}~\bibnamefont {Jia}},\ and\
  \bibinfo {author} {\bibfnamefont {M.~Z.}\ \bibnamefont {Hasan}},\ }\bibfield
  {title} {\bibinfo {title} {{D}iscovery of a {W}eyl fermion semimetal and
  topological {F}ermi arcs},\ }\href {https://doi.org/10.1126/science.aaa9297}
  {\bibfield  {journal} {\bibinfo  {journal} {Science}\ }\textbf {\bibinfo
  {volume} {349}},\ \bibinfo {pages} {613} (\bibinfo {year}
  {2015}{\natexlab{b}})}\BibitemShut {NoStop}%
\bibitem [{\citenamefont {Xu}\ \emph {et~al.}(2015{\natexlab{c}})\citenamefont
  {Xu}, \citenamefont {{A}lidoust}, \citenamefont {Belopolski}, \citenamefont
  {Yuan}, \citenamefont {Bian}, \citenamefont {Chang}, \citenamefont {Zheng},
  \citenamefont {Strocov}, \citenamefont {Sanchez}, \citenamefont {Chang},
  \citenamefont {Zhang}, \citenamefont {Mou}, \citenamefont {{W}u},
  \citenamefont {Huang}, \citenamefont {Lee}, \citenamefont {Huang},
  \citenamefont {{W}ang}, \citenamefont {Bansil}, \citenamefont {Jeng},
  \citenamefont {Neupert}, \citenamefont {Kaminski}, \citenamefont {Lin},
  \citenamefont {Jia},\ and\ \citenamefont {Zahid~Hasan}}]{Xu2015NbAs}%
  \BibitemOpen
  \bibfield  {author} {\bibinfo {author} {\bibfnamefont {S.-Y.}\ \bibnamefont
  {Xu}}, \bibinfo {author} {\bibfnamefont {N.}~\bibnamefont {{A}lidoust}},
  \bibinfo {author} {\bibfnamefont {I.}~\bibnamefont {Belopolski}}, \bibinfo
  {author} {\bibfnamefont {Z.}~\bibnamefont {Yuan}}, \bibinfo {author}
  {\bibfnamefont {G.}~\bibnamefont {Bian}}, \bibinfo {author} {\bibfnamefont
  {T.-R.}\ \bibnamefont {Chang}}, \bibinfo {author} {\bibfnamefont
  {H.}~\bibnamefont {Zheng}}, \bibinfo {author} {\bibfnamefont {V.~N.}\
  \bibnamefont {Strocov}}, \bibinfo {author} {\bibfnamefont {D.~S.}\
  \bibnamefont {Sanchez}}, \bibinfo {author} {\bibfnamefont {G.}~\bibnamefont
  {Chang}}, \bibinfo {author} {\bibfnamefont {C.}~\bibnamefont {Zhang}},
  \bibinfo {author} {\bibfnamefont {D.}~\bibnamefont {Mou}}, \bibinfo {author}
  {\bibfnamefont {Y.}~\bibnamefont {{W}u}}, \bibinfo {author} {\bibfnamefont
  {L.}~\bibnamefont {Huang}}, \bibinfo {author} {\bibfnamefont {C.-C.}\
  \bibnamefont {Lee}}, \bibinfo {author} {\bibfnamefont {S.-M.}\ \bibnamefont
  {Huang}}, \bibinfo {author} {\bibfnamefont {B.}~\bibnamefont {{W}ang}},
  \bibinfo {author} {\bibfnamefont {A.}~\bibnamefont {Bansil}}, \bibinfo
  {author} {\bibfnamefont {H.-T.}\ \bibnamefont {Jeng}}, \bibinfo {author}
  {\bibfnamefont {T.}~\bibnamefont {Neupert}}, \bibinfo {author} {\bibfnamefont
  {A.}~\bibnamefont {Kaminski}}, \bibinfo {author} {\bibfnamefont
  {H.}~\bibnamefont {Lin}}, \bibinfo {author} {\bibfnamefont {S.}~\bibnamefont
  {Jia}},\ and\ \bibinfo {author} {\bibfnamefont {M.}~\bibnamefont
  {Zahid~Hasan}},\ }\bibfield  {title} {\bibinfo {title} {{D}iscovery of a
  {W}eyl fermion state with {F}ermi arcs in niobium arsenide},\ }\href
  {http://dx.doi.org/10.1038/nphys3437} {\bibfield  {journal} {\bibinfo
  {journal} {Nat. Phys.}\ }\textbf {\bibinfo {volume} {11}},\ \bibinfo {pages}
  {748} (\bibinfo {year} {2015}{\natexlab{c}})}\BibitemShut {NoStop}%
\bibitem [{\citenamefont {Hirayama}\ \emph {et~al.}(2015)\citenamefont
  {Hirayama}, \citenamefont {Okugawa}, \citenamefont {Ishibashi}, \citenamefont
  {Murakami},\ and\ \citenamefont {Miyake}}]{Hirayama2015}%
  \BibitemOpen
  \bibfield  {author} {\bibinfo {author} {\bibfnamefont {M.}~\bibnamefont
  {Hirayama}}, \bibinfo {author} {\bibfnamefont {R.}~\bibnamefont {Okugawa}},
  \bibinfo {author} {\bibfnamefont {S.}~\bibnamefont {Ishibashi}}, \bibinfo
  {author} {\bibfnamefont {S.}~\bibnamefont {Murakami}},\ and\ \bibinfo
  {author} {\bibfnamefont {T.}~\bibnamefont {Miyake}},\ }\bibfield  {title}
  {\bibinfo {title} {Weyl {N}ode and {S}pin {T}exture in {T}rigonal {T}ellurium
  and {S}elenium},\ }\href {https://doi.org/10.1103/PhysRevLett.114.206401}
  {\bibfield  {journal} {\bibinfo  {journal} {Phys. Rev. Lett.}\ }\textbf
  {\bibinfo {volume} {114}},\ \bibinfo {pages} {206401} (\bibinfo {year}
  {2015})}\BibitemShut {NoStop}%
\bibitem [{\citenamefont {{Ruan}}\ \emph {et~al.}(2016)\citenamefont {{Ruan}},
  \citenamefont {{Jian}}, \citenamefont {{Yao}}, \citenamefont {{Zhang}},
  \citenamefont {{Zhang}},\ and\ \citenamefont {{Xing}}}]{Ruan2016HgTe}%
  \BibitemOpen
  \bibfield  {author} {\bibinfo {author} {\bibfnamefont {J.}~\bibnamefont
  {{Ruan}}}, \bibinfo {author} {\bibfnamefont {S.-K.}\ \bibnamefont {{Jian}}},
  \bibinfo {author} {\bibfnamefont {H.}~\bibnamefont {{Yao}}}, \bibinfo
  {author} {\bibfnamefont {H.}~\bibnamefont {{Zhang}}}, \bibinfo {author}
  {\bibfnamefont {S.-C.}\ \bibnamefont {{Zhang}}},\ and\ \bibinfo {author}
  {\bibfnamefont {D.}~\bibnamefont {{Xing}}},\ }\bibfield  {title} {\bibinfo
  {title} {{Symmetry-protected ideal {W}eyl semimetal in {H}g{T}e-class
  materials}},\ }\href {https://doi.org/10.1038/ncomms11136} {\bibfield
  {journal} {\bibinfo  {journal} {Nat. Commun.}\ }\textbf {\bibinfo {volume}
  {7}},\ \bibinfo {pages} {11136} (\bibinfo {year} {2016})}\BibitemShut
  {NoStop}%
\bibitem [{\citenamefont {Ruan}\ \emph {et~al.}(2016)\citenamefont {Ruan},
  \citenamefont {Jian}, \citenamefont {Zhang}, \citenamefont {Yao},
  \citenamefont {Zhang}, \citenamefont {Zhang},\ and\ \citenamefont
  {Xing}}]{Ruan2016Chalk}%
  \BibitemOpen
  \bibfield  {author} {\bibinfo {author} {\bibfnamefont {J.}~\bibnamefont
  {Ruan}}, \bibinfo {author} {\bibfnamefont {S.-K.}\ \bibnamefont {Jian}},
  \bibinfo {author} {\bibfnamefont {D.}~\bibnamefont {Zhang}}, \bibinfo
  {author} {\bibfnamefont {H.}~\bibnamefont {Yao}}, \bibinfo {author}
  {\bibfnamefont {H.}~\bibnamefont {Zhang}}, \bibinfo {author} {\bibfnamefont
  {S.-C.}\ \bibnamefont {Zhang}},\ and\ \bibinfo {author} {\bibfnamefont
  {D.}~\bibnamefont {Xing}},\ }\bibfield  {title} {\bibinfo {title} {Ideal
  {W}eyl semimetals in the {C}halcopyrites {C}u{T}l{S}e$_2$,
  {A}g{T}l{T}e$_{2}$, {A}u{T}l{T}e$_{2}$, and {Z}n{P}b{A}s$_{2}$},\ }\href
  {https://doi.org/10.1103/PhysRevLett.116.226801} {\bibfield  {journal}
  {\bibinfo  {journal} {Phys. Rev. Lett.}\ }\textbf {\bibinfo {volume} {116}},\
  \bibinfo {pages} {226801} (\bibinfo {year} {2016})}\BibitemShut {NoStop}%
\bibitem [{\citenamefont {Lee}\ \emph {et~al.}(2015)\citenamefont {Lee},
  \citenamefont {Xu}, \citenamefont {Huang}, \citenamefont {Sanchez},
  \citenamefont {Belopolski}, \citenamefont {Chang}, \citenamefont {Bian},
  \citenamefont {Alidoust}, \citenamefont {Zheng}, \citenamefont {Neupane},
  \citenamefont {Wang}, \citenamefont {Bansil}, \citenamefont {Hasan},\ and\
  \citenamefont {Lin}}]{Lee2015}%
  \BibitemOpen
  \bibfield  {author} {\bibinfo {author} {\bibfnamefont {C.-C.}\ \bibnamefont
  {Lee}}, \bibinfo {author} {\bibfnamefont {S.-Y.}\ \bibnamefont {Xu}},
  \bibinfo {author} {\bibfnamefont {S.-M.}\ \bibnamefont {Huang}}, \bibinfo
  {author} {\bibfnamefont {D.~S.}\ \bibnamefont {Sanchez}}, \bibinfo {author}
  {\bibfnamefont {I.}~\bibnamefont {Belopolski}}, \bibinfo {author}
  {\bibfnamefont {G.}~\bibnamefont {Chang}}, \bibinfo {author} {\bibfnamefont
  {G.}~\bibnamefont {Bian}}, \bibinfo {author} {\bibfnamefont {N.}~\bibnamefont
  {Alidoust}}, \bibinfo {author} {\bibfnamefont {H.}~\bibnamefont {Zheng}},
  \bibinfo {author} {\bibfnamefont {M.}~\bibnamefont {Neupane}}, \bibinfo
  {author} {\bibfnamefont {B.}~\bibnamefont {Wang}}, \bibinfo {author}
  {\bibfnamefont {A.}~\bibnamefont {Bansil}}, \bibinfo {author} {\bibfnamefont
  {M.~Z.}\ \bibnamefont {Hasan}},\ and\ \bibinfo {author} {\bibfnamefont
  {H.}~\bibnamefont {Lin}},\ }\bibfield  {title} {\bibinfo {title} {Fermi
  surface interconnectivity and topology in {W}eyl fermion semimetals {T}a{A}s,
  {T}a{P}, {N}b{A}s, and {N}b{P}},\ }\href
  {https://doi.org/10.1103/PhysRevB.92.235104} {\bibfield  {journal} {\bibinfo
  {journal} {Phys. Rev. B}\ }\textbf {\bibinfo {volume} {92}},\ \bibinfo
  {pages} {235104} (\bibinfo {year} {2015})}\BibitemShut {NoStop}%
\bibitem [{\citenamefont {{A}rnold}\ \emph {et~al.}(2016)\citenamefont
  {{A}rnold}, \citenamefont {Shekhar}, \citenamefont {{W}u}, \citenamefont
  {Sun}, \citenamefont {dos Reis}, \citenamefont {Kumar}, \citenamefont
  {Naumann}, \citenamefont {{A}jeesh}, \citenamefont {Schmidt}, \citenamefont
  {Grushin}, \citenamefont {Bardarson}, \citenamefont {Baenitz}, \citenamefont
  {Sokolov}, \citenamefont {Borrmann}, \citenamefont {Nicklas}, \citenamefont
  {Felser}, \citenamefont {Hassinger},\ and\ \citenamefont {Yan}}]{Arnold2016}%
  \BibitemOpen
  \bibfield  {author} {\bibinfo {author} {\bibfnamefont {F.}~\bibnamefont
  {{A}rnold}}, \bibinfo {author} {\bibfnamefont {C.}~\bibnamefont {Shekhar}},
  \bibinfo {author} {\bibfnamefont {S.-C.}\ \bibnamefont {{W}u}}, \bibinfo
  {author} {\bibfnamefont {Y.}~\bibnamefont {Sun}}, \bibinfo {author}
  {\bibfnamefont {R.~D.}\ \bibnamefont {dos Reis}}, \bibinfo {author}
  {\bibfnamefont {N.}~\bibnamefont {Kumar}}, \bibinfo {author} {\bibfnamefont
  {M.}~\bibnamefont {Naumann}}, \bibinfo {author} {\bibfnamefont {M.~O.}\
  \bibnamefont {{A}jeesh}}, \bibinfo {author} {\bibfnamefont {M.}~\bibnamefont
  {Schmidt}}, \bibinfo {author} {\bibfnamefont {A.~G.}\ \bibnamefont
  {Grushin}}, \bibinfo {author} {\bibfnamefont {J.~H.}\ \bibnamefont
  {Bardarson}}, \bibinfo {author} {\bibfnamefont {M.}~\bibnamefont {Baenitz}},
  \bibinfo {author} {\bibfnamefont {D.}~\bibnamefont {Sokolov}}, \bibinfo
  {author} {\bibfnamefont {H.}~\bibnamefont {Borrmann}}, \bibinfo {author}
  {\bibfnamefont {M.}~\bibnamefont {Nicklas}}, \bibinfo {author} {\bibfnamefont
  {C.}~\bibnamefont {Felser}}, \bibinfo {author} {\bibfnamefont
  {E.}~\bibnamefont {Hassinger}},\ and\ \bibinfo {author} {\bibfnamefont
  {B.}~\bibnamefont {Yan}},\ }\bibfield  {title} {\bibinfo {title} {Negative
  magnetoresistance without well-defined chirality in the {W}eyl semimetal
  {T}a{P}},\ }\href {http://dx.doi.org/10.1038/ncomms11615} {\bibfield
  {journal} {\bibinfo  {journal} {Nat. Commun.}\ }\textbf {\bibinfo {volume}
  {7}},\ \bibinfo {pages} {11615} (\bibinfo {year} {2016})}\BibitemShut
  {NoStop}%
\bibitem [{\citenamefont {Hu}\ \emph {et~al.}(2016)\citenamefont {Hu},
  \citenamefont {Liu}, \citenamefont {Graf}, \citenamefont {Radmanesh},
  \citenamefont {{A}dams}, \citenamefont {Chuang}, \citenamefont {{W}ang},
  \citenamefont {Chiorescu}, \citenamefont {{W}ei}, \citenamefont {Spinu},\
  and\ \citenamefont {Mao}}]{Hu2016}%
  \BibitemOpen
  \bibfield  {author} {\bibinfo {author} {\bibfnamefont {J.}~\bibnamefont
  {Hu}}, \bibinfo {author} {\bibfnamefont {J.~Y.}\ \bibnamefont {Liu}},
  \bibinfo {author} {\bibfnamefont {D.}~\bibnamefont {Graf}}, \bibinfo {author}
  {\bibfnamefont {S.~M.~A.}\ \bibnamefont {Radmanesh}}, \bibinfo {author}
  {\bibfnamefont {D.~J.}\ \bibnamefont {{A}dams}}, \bibinfo {author}
  {\bibfnamefont {A.}~\bibnamefont {Chuang}}, \bibinfo {author} {\bibfnamefont
  {Y.}~\bibnamefont {{W}ang}}, \bibinfo {author} {\bibfnamefont
  {I.}~\bibnamefont {Chiorescu}}, \bibinfo {author} {\bibfnamefont
  {J.}~\bibnamefont {{W}ei}}, \bibinfo {author} {\bibfnamefont
  {L.}~\bibnamefont {Spinu}},\ and\ \bibinfo {author} {\bibfnamefont {Z.~Q.}\
  \bibnamefont {Mao}},\ }\bibfield  {title} {\bibinfo {title} {$\pi$-{B}erry
  phase and {Z}eeman splitting of {W}eyl semimetal {T}a{P}},\ }\href
  {http://dx.doi.org/10.1038/srep18674} {\bibfield  {journal} {\bibinfo
  {journal} {Sci. Rep.}\ }\textbf {\bibinfo {volume} {6}},\ \bibinfo {pages}
  {18674} (\bibinfo {year} {2016})}\BibitemShut {NoStop}%
\bibitem [{\citenamefont {Klotz}\ \emph {et~al.}(2016)\citenamefont {Klotz},
  \citenamefont {{W}u}, \citenamefont {Shekhar}, \citenamefont {Sun},
  \citenamefont {Schmidt}, \citenamefont {Nicklas}, \citenamefont {Baenitz},
  \citenamefont {Uhlarz}, \citenamefont {{W}osnitza}, \citenamefont {Felser},\
  and\ \citenamefont {Yan}}]{Klotz2016}%
  \BibitemOpen
  \bibfield  {author} {\bibinfo {author} {\bibfnamefont {J.}~\bibnamefont
  {Klotz}}, \bibinfo {author} {\bibfnamefont {S.-C.}\ \bibnamefont {{W}u}},
  \bibinfo {author} {\bibfnamefont {C.}~\bibnamefont {Shekhar}}, \bibinfo
  {author} {\bibfnamefont {Y.}~\bibnamefont {Sun}}, \bibinfo {author}
  {\bibfnamefont {M.}~\bibnamefont {Schmidt}}, \bibinfo {author} {\bibfnamefont
  {M.}~\bibnamefont {Nicklas}}, \bibinfo {author} {\bibfnamefont
  {M.}~\bibnamefont {Baenitz}}, \bibinfo {author} {\bibfnamefont
  {M.}~\bibnamefont {Uhlarz}}, \bibinfo {author} {\bibfnamefont
  {J.}~\bibnamefont {{W}osnitza}}, \bibinfo {author} {\bibfnamefont
  {C.}~\bibnamefont {Felser}},\ and\ \bibinfo {author} {\bibfnamefont
  {B.}~\bibnamefont {Yan}},\ }\bibfield  {title} {\bibinfo {title} {Quantum
  oscillations and the {F}ermi surface topology of the {W}eyl semimetal
  {N}b{P}},\ }\href {https://doi.org/10.1103/PhysRevB.93.121105} {\bibfield
  {journal} {\bibinfo  {journal} {Phys. Rev. B}\ }\textbf {\bibinfo {volume}
  {93}},\ \bibinfo {pages} {121105} (\bibinfo {year} {2016})}\BibitemShut
  {NoStop}%
\bibitem [{\citenamefont {Grassano}\ \emph {et~al.}(2018)\citenamefont
  {Grassano}, \citenamefont {Pulci}, \citenamefont {Conte},\ and\ \citenamefont
  {Bechstedt}}]{Grassano2018}%
  \BibitemOpen
  \bibfield  {author} {\bibinfo {author} {\bibfnamefont {D.}~\bibnamefont
  {Grassano}}, \bibinfo {author} {\bibfnamefont {O.}~\bibnamefont {Pulci}},
  \bibinfo {author} {\bibfnamefont {A.~M.}\ \bibnamefont {Conte}},\ and\
  \bibinfo {author} {\bibfnamefont {F.}~\bibnamefont {Bechstedt}},\ }\bibfield
  {title} {\bibinfo {title} {Validity of {W}eyl fermion picture for transition
  metals monopnictides {T}a{A}s, {T}a{P}, {N}b{A}s, and {N}b{P} from ab initio
  studies},\ }\href {https://www.nature.com/articles/s41598-018-21465-z}
  {\bibfield  {journal} {\bibinfo  {journal} {Sci. Rep.}\ }\textbf {\bibinfo
  {volume} {8}},\ \bibinfo {pages} {3534} (\bibinfo {year} {2018})}\BibitemShut
  {NoStop}%
\bibitem [{\citenamefont {Sadhukhan}\ \emph {et~al.}(2020)\citenamefont
  {Sadhukhan}, \citenamefont {Politano},\ and\ \citenamefont
  {Agarwal}}]{Sadhukhan2020}%
  \BibitemOpen
  \bibfield  {author} {\bibinfo {author} {\bibfnamefont {K.}~\bibnamefont
  {Sadhukhan}}, \bibinfo {author} {\bibfnamefont {A.}~\bibnamefont
  {Politano}},\ and\ \bibinfo {author} {\bibfnamefont {A.}~\bibnamefont
  {Agarwal}},\ }\bibfield  {title} {\bibinfo {title} {Novel {U}ndamped
  {G}apless {P}lasmon {M}ode in a {T}ilted {T}ype-{II} {D}irac {S}emimetal},\
  }\href {https://doi.org/10.1103/PhysRevLett.124.046803} {\bibfield  {journal}
  {\bibinfo  {journal} {Phys. Rev. Lett.}\ }\textbf {\bibinfo {volume} {124}},\
  \bibinfo {pages} {046803} (\bibinfo {year} {2020})}\BibitemShut {NoStop}%
\bibitem [{\citenamefont {Chan}\ \emph {et~al.}(2017)\citenamefont {Chan},
  \citenamefont {Lindner}, \citenamefont {Refael},\ and\ \citenamefont
  {Lee}}]{Chan2017}%
  \BibitemOpen
  \bibfield  {author} {\bibinfo {author} {\bibfnamefont {C.-K.}\ \bibnamefont
  {Chan}}, \bibinfo {author} {\bibfnamefont {N.~H.}\ \bibnamefont {Lindner}},
  \bibinfo {author} {\bibfnamefont {G.}~\bibnamefont {Refael}},\ and\ \bibinfo
  {author} {\bibfnamefont {P.~A.}\ \bibnamefont {Lee}},\ }\bibfield  {title}
  {\bibinfo {title} {Photocurrents in {W}eyl semimetals},\ }\href
  {https://doi.org/10.1103/PhysRevB.95.041104} {\bibfield  {journal} {\bibinfo
  {journal} {Phys. Rev. B}\ }\textbf {\bibinfo {volume} {95}},\ \bibinfo
  {pages} {041104} (\bibinfo {year} {2017})}\BibitemShut {NoStop}%
\bibitem [{\citenamefont {Golub}\ and\ \citenamefont
  {Ivchenko}(2018)}]{Golub2018}%
  \BibitemOpen
  \bibfield  {author} {\bibinfo {author} {\bibfnamefont {L.~E.}\ \bibnamefont
  {Golub}}\ and\ \bibinfo {author} {\bibfnamefont {E.~L.}\ \bibnamefont
  {Ivchenko}},\ }\bibfield  {title} {\bibinfo {title} {Circular and
  magnetoinduced photocurrents in {W}eyl semimetals},\ }\href
  {https://doi.org/10.1103/PhysRevB.98.075305} {\bibfield  {journal} {\bibinfo
  {journal} {Phys. Rev. B}\ }\textbf {\bibinfo {volume} {98}},\ \bibinfo
  {pages} {075305} (\bibinfo {year} {2018})}\BibitemShut {NoStop}%
\bibitem [{\citenamefont {Afanasiev}\ \emph {et~al.}(2019)\citenamefont
  {Afanasiev}, \citenamefont {Greshnov},\ and\ \citenamefont
  {Svintsov}}]{Afanasiev2019}%
  \BibitemOpen
  \bibfield  {author} {\bibinfo {author} {\bibfnamefont {A.~N.}\ \bibnamefont
  {Afanasiev}}, \bibinfo {author} {\bibfnamefont {A.~A.}\ \bibnamefont
  {Greshnov}},\ and\ \bibinfo {author} {\bibfnamefont {D.}~\bibnamefont
  {Svintsov}},\ }\bibfield  {title} {\bibinfo {title} {Relativistic suppression
  of {A}uger recombination in {W}eyl semimetals},\ }\href
  {https://doi.org/10.1103/PhysRevB.99.115202} {\bibfield  {journal} {\bibinfo
  {journal} {Phys. Rev. B}\ }\textbf {\bibinfo {volume} {99}},\ \bibinfo
  {pages} {115202} (\bibinfo {year} {2019})}\BibitemShut {NoStop}%
\bibitem [{HgT()}]{HgTe_Fermi_level}%
  \BibitemOpen
  \href@noop {} {\ }\bibinfo {note} {In intrinsic WSMs of HgTe family the Fermi
  level is exactly at the Weyl crossing point.}\BibitemShut {Stop}%
\bibitem [{TaA()}]{TaAs_Not_Semimetal}%
  \BibitemOpen
  \href@noop {} {\ }\bibinfo {note} {According to electronic band theory,
  materials of the TaAs family are not technically semimetals, since the Fermi
  level is never at the Weyl crossing point of $W_1$ and $W_2$ node
  groups.}\BibitemShut {Stop}%
\bibitem [{\citenamefont {Pines}(1956)}]{Pines1956}%
  \BibitemOpen
  \bibfield  {author} {\bibinfo {author} {\bibfnamefont {D.}~\bibnamefont
  {Pines}},\ }\bibfield  {title} {\bibinfo {title} {Electron interaction in
  solids},\ }\href {https://doi.org/10.1139/p56-154} {\bibfield  {journal}
  {\bibinfo  {journal} {Canadian Journal of Physics}\ }\textbf {\bibinfo
  {volume} {34}},\ \bibinfo {pages} {1379} (\bibinfo {year} {1956})},\ \Eprint
  {https://arxiv.org/abs/https://doi.org/10.1139/p56-154}
  {https://doi.org/10.1139/p56-154} \BibitemShut {NoStop}%
\bibitem [{\citenamefont {Giuliani}\ and\ \citenamefont
  {Vignale}(2005)}]{Giuliani2005}%
  \BibitemOpen
  \bibfield  {author} {\bibinfo {author} {\bibfnamefont {G.}~\bibnamefont
  {Giuliani}}\ and\ \bibinfo {author} {\bibfnamefont {G.}~\bibnamefont
  {Vignale}},\ }\href {https://doi.org/10.1017/CBO9780511619915} {\emph
  {\bibinfo {title} {Quantum {T}heory of the {E}lectron {L}iquid}}}\ (\bibinfo
  {publisher} {Cambridge University Press},\ \bibinfo {year}
  {2005})\BibitemShut {NoStop}%
\bibitem [{\citenamefont {Abrikosov}\ and\ \citenamefont
  {Beneslavskii}(1971)}]{AbrikosovBeneslavskii1971_Cutoff}%
  \BibitemOpen
  \bibfield  {author} {\bibinfo {author} {\bibfnamefont {A.~A.}\ \bibnamefont
  {Abrikosov}}\ and\ \bibinfo {author} {\bibfnamefont {S.~D.}\ \bibnamefont
  {Beneslavskii}},\ }\bibfield  {title} {\bibinfo {title} {Some properties of
  gapless semiconductors of the second kind},\ }\href
  {https://doi.org/10.1007/BF00629569} {\bibfield  {journal} {\bibinfo
  {journal} {J. Low Temp. Phys.}\ }\textbf {\bibinfo {volume} {5}},\ \bibinfo
  {pages} {141} (\bibinfo {year} {1971})}\BibitemShut {NoStop}%
\bibitem [{\citenamefont {Agarwal}\ \emph {et~al.}(2014)\citenamefont
  {Agarwal}, \citenamefont {Polini}, \citenamefont {Vignale},\ and\
  \citenamefont {Flatt\'e}}]{Agarwal2014}%
  \BibitemOpen
  \bibfield  {author} {\bibinfo {author} {\bibfnamefont {A.}~\bibnamefont
  {Agarwal}}, \bibinfo {author} {\bibfnamefont {M.}~\bibnamefont {Polini}},
  \bibinfo {author} {\bibfnamefont {G.}~\bibnamefont {Vignale}},\ and\ \bibinfo
  {author} {\bibfnamefont {M.~E.}\ \bibnamefont {Flatt\'e}},\ }\bibfield
  {title} {\bibinfo {title} {Long-lived spin plasmons in a spin-polarized
  two-dimensional electron gas},\ }\href
  {https://doi.org/10.1103/PhysRevB.90.155409} {\bibfield  {journal} {\bibinfo
  {journal} {Phys. Rev. B}\ }\textbf {\bibinfo {volume} {90}},\ \bibinfo
  {pages} {155409} (\bibinfo {year} {2014})}\BibitemShut {NoStop}%
\bibitem [{\citenamefont {Falkovsky}(2011)}]{Falkovsky2011}%
  \BibitemOpen
  \bibfield  {author} {\bibinfo {author} {\bibfnamefont {L.~A.}\ \bibnamefont
  {Falkovsky}},\ }\bibfield  {title} {\bibinfo {title} {Optics of
  semiconductors with a linear electron spectrum},\ }\href
  {https://doi.org/10.1063/1.3615524} {\bibfield  {journal} {\bibinfo
  {journal} {Low Temp. Phys.}\ }\textbf {\bibinfo {volume} {37}},\ \bibinfo
  {pages} {480} (\bibinfo {year} {2011})}\BibitemShut {NoStop}%
\bibitem [{\citenamefont {Hwang}\ and\ \citenamefont
  {Das~Sarma}(2007)}]{Hwang2007}%
  \BibitemOpen
  \bibfield  {author} {\bibinfo {author} {\bibfnamefont {E.~H.}\ \bibnamefont
  {Hwang}}\ and\ \bibinfo {author} {\bibfnamefont {S.}~\bibnamefont
  {Das~Sarma}},\ }\bibfield  {title} {\bibinfo {title} {Dielectric function,
  screening, and plasmons in two-dimensional graphene},\ }\href
  {https://doi.org/10.1103/PhysRevB.75.205418} {\bibfield  {journal} {\bibinfo
  {journal} {Phys. Rev. B}\ }\textbf {\bibinfo {volume} {75}},\ \bibinfo
  {pages} {205418} (\bibinfo {year} {2007})}\BibitemShut {NoStop}%
\bibitem [{\citenamefont {Ashoori}\ \emph {et~al.}(1992)\citenamefont
  {Ashoori}, \citenamefont {Stormer}, \citenamefont {Pfeiffer}, \citenamefont
  {Baldwin},\ and\ \citenamefont {West}}]{Ashoori1992}%
  \BibitemOpen
  \bibfield  {author} {\bibinfo {author} {\bibfnamefont {R.~C.}\ \bibnamefont
  {Ashoori}}, \bibinfo {author} {\bibfnamefont {H.~L.}\ \bibnamefont
  {Stormer}}, \bibinfo {author} {\bibfnamefont {L.~N.}\ \bibnamefont
  {Pfeiffer}}, \bibinfo {author} {\bibfnamefont {K.~W.}\ \bibnamefont
  {Baldwin}},\ and\ \bibinfo {author} {\bibfnamefont {K.}~\bibnamefont
  {West}},\ }\bibfield  {title} {\bibinfo {title} {Edge magnetoplasmons in the
  time domain},\ }\href {https://doi.org/10.1103/PhysRevB.45.3894} {\bibfield
  {journal} {\bibinfo  {journal} {Phys. Rev. B}\ }\textbf {\bibinfo {volume}
  {45}},\ \bibinfo {pages} {3894} (\bibinfo {year} {1992})}\BibitemShut
  {NoStop}%
\bibitem [{\citenamefont {Kumada}\ \emph {et~al.}(2014)\citenamefont {Kumada},
  \citenamefont {Roulleau}, \citenamefont {Roche}, \citenamefont {Hashisaka},
  \citenamefont {Hibino}, \citenamefont {Petkovi\ifmmode~\acute{c}\else
  \'{c}\fi{}},\ and\ \citenamefont {Glattli}}]{Kumada2014}%
  \BibitemOpen
  \bibfield  {author} {\bibinfo {author} {\bibfnamefont {N.}~\bibnamefont
  {Kumada}}, \bibinfo {author} {\bibfnamefont {P.}~\bibnamefont {Roulleau}},
  \bibinfo {author} {\bibfnamefont {B.}~\bibnamefont {Roche}}, \bibinfo
  {author} {\bibfnamefont {M.}~\bibnamefont {Hashisaka}}, \bibinfo {author}
  {\bibfnamefont {H.}~\bibnamefont {Hibino}}, \bibinfo {author} {\bibfnamefont
  {I.}~\bibnamefont {Petkovi\ifmmode~\acute{c}\else \'{c}\fi{}}},\ and\
  \bibinfo {author} {\bibfnamefont {D.~C.}\ \bibnamefont {Glattli}},\
  }\bibfield  {title} {\bibinfo {title} {Resonant edge magnetoplasmons and
  their decay in graphene},\ }\href
  {https://doi.org/10.1103/PhysRevLett.113.266601} {\bibfield  {journal}
  {\bibinfo  {journal} {Phys. Rev. Lett.}\ }\textbf {\bibinfo {volume} {113}},\
  \bibinfo {pages} {266601} (\bibinfo {year} {2014})}\BibitemShut {NoStop}%
\bibitem [{\citenamefont {Lifshitz}\ and\ \citenamefont
  {Pitaevski\u{i}}(1981)}]{Landau_V10}%
  \BibitemOpen
  \bibfield  {author} {\bibinfo {author} {\bibfnamefont {E.~M.}\ \bibnamefont
  {Lifshitz}}\ and\ \bibinfo {author} {\bibfnamefont {L.~P.}\ \bibnamefont
  {Pitaevski\u{i}}},\ }\href
  {https://books.google.ru/books?id=DTHxPDfV0fQC&dq=lifshitz+pitaevskii&lr=&hl=ru&source=gbs_navlinks_s}
  {\emph {\bibinfo {title} {Landau and Lifshitz Course of Theoretical Physics:
  Physical Kinetics}}},\ Vol.~\bibinfo {volume} {10}\ (\bibinfo  {publisher}
  {Pergamon Press},\ \bibinfo {year} {1981})\BibitemShut {NoStop}%
\bibitem [{\citenamefont {Badalyan}\ \emph {et~al.}(2013)\citenamefont
  {Badalyan}, \citenamefont {Matos-Abiague}, \citenamefont {Fabian},
  \citenamefont {Vignale},\ and\ \citenamefont {Peeters}}]{Badalyan2013}%
  \BibitemOpen
  \bibfield  {author} {\bibinfo {author} {\bibfnamefont {S.~M.}\ \bibnamefont
  {Badalyan}}, \bibinfo {author} {\bibfnamefont {A.}~\bibnamefont
  {Matos-Abiague}}, \bibinfo {author} {\bibfnamefont {J.}~\bibnamefont
  {Fabian}}, \bibinfo {author} {\bibfnamefont {G.}~\bibnamefont {Vignale}},\
  and\ \bibinfo {author} {\bibfnamefont {F.~M.}\ \bibnamefont {Peeters}},\
  }\bibfield  {title} {\bibinfo {title} {Spin-orbit-interaction induced
  singularity of the charge density relaxation propagator},\ }\href
  {https://doi.org/10.1103/PhysRevB.88.195402} {\bibfield  {journal} {\bibinfo
  {journal} {Phys. Rev. B}\ }\textbf {\bibinfo {volume} {88}},\ \bibinfo
  {pages} {195402} (\bibinfo {year} {2013})}\BibitemShut {NoStop}%
\bibitem [{\citenamefont {Cao}\ \emph {et~al.}(2018)\citenamefont {Cao},
  \citenamefont {Fatemi}, \citenamefont {Fang}, \citenamefont {Watanabe},
  \citenamefont {Taniguchi}, \citenamefont {Kaxiras},\ and\ \citenamefont
  {Jarillo-Herrero}}]{Cao2018}%
  \BibitemOpen
  \bibfield  {author} {\bibinfo {author} {\bibfnamefont {Y.}~\bibnamefont
  {Cao}}, \bibinfo {author} {\bibfnamefont {V.}~\bibnamefont {Fatemi}},
  \bibinfo {author} {\bibfnamefont {S.}~\bibnamefont {Fang}}, \bibinfo {author}
  {\bibfnamefont {K.}~\bibnamefont {Watanabe}}, \bibinfo {author}
  {\bibfnamefont {T.}~\bibnamefont {Taniguchi}}, \bibinfo {author}
  {\bibfnamefont {E.}~\bibnamefont {Kaxiras}},\ and\ \bibinfo {author}
  {\bibfnamefont {P.}~\bibnamefont {Jarillo-Herrero}},\ }\bibfield  {title}
  {\bibinfo {title} {Unconventional superconductivity in magic-angle graphene
  superlattices},\ }\href {https://doi.org/10.1038/nature26160} {\bibfield
  {journal} {\bibinfo  {journal} {Nature}\ }\textbf {\bibinfo {volume} {556}},\
  \bibinfo {pages} {43} (\bibinfo {year} {2018})}\BibitemShut {NoStop}%
\bibitem [{\citenamefont {Sharma}\ \emph {et~al.}(2020)\citenamefont {Sharma},
  \citenamefont {Trushin}, \citenamefont {Sushkov}, \citenamefont {Vignale},\
  and\ \citenamefont {Adam}}]{Sharma2020}%
  \BibitemOpen
  \bibfield  {author} {\bibinfo {author} {\bibfnamefont {G.}~\bibnamefont
  {Sharma}}, \bibinfo {author} {\bibfnamefont {M.}~\bibnamefont {Trushin}},
  \bibinfo {author} {\bibfnamefont {O.~P.}\ \bibnamefont {Sushkov}}, \bibinfo
  {author} {\bibfnamefont {G.}~\bibnamefont {Vignale}},\ and\ \bibinfo {author}
  {\bibfnamefont {S.}~\bibnamefont {Adam}},\ }\bibfield  {title} {\bibinfo
  {title} {Superconductivity from collective excitations in magic-angle twisted
  bilayer graphene},\ }\href {https://doi.org/10.1103/PhysRevResearch.2.022040}
  {\bibfield  {journal} {\bibinfo  {journal} {Phys. Rev. Research}\ }\textbf
  {\bibinfo {volume} {2}},\ \bibinfo {pages} {022040} (\bibinfo {year}
  {2020})}\BibitemShut {NoStop}%
\bibitem [{\citenamefont {Fröhlich}(1968)}]{Frohlich1968}%
  \BibitemOpen
  \bibfield  {author} {\bibinfo {author} {\bibfnamefont {H.}~\bibnamefont
  {Fröhlich}},\ }\bibfield  {title} {\bibinfo {title} {Superconductivity in
  metals with incomplete inner shells},\ }\href
  {https://doi.org/10.1088/0022-3719/1/2/131} {\bibfield  {journal} {\bibinfo
  {journal} {Journal of Physics C: Solid State Physics}\ }\textbf {\bibinfo
  {volume} {1}},\ \bibinfo {pages} {544} (\bibinfo {year} {1968})}\BibitemShut
  {NoStop}%
\bibitem [{\citenamefont {Canright}\ and\ \citenamefont
  {Vignale}(1989)}]{Canright1989}%
  \BibitemOpen
  \bibfield  {author} {\bibinfo {author} {\bibfnamefont {G.~S.}\ \bibnamefont
  {Canright}}\ and\ \bibinfo {author} {\bibfnamefont {G.}~\bibnamefont
  {Vignale}},\ }\bibfield  {title} {\bibinfo {title} {Superconductivity and
  acoustic plasmons in the two-dimensional electron gas},\ }\href
  {https://doi.org/10.1103/PhysRevB.39.2740} {\bibfield  {journal} {\bibinfo
  {journal} {Phys. Rev. B}\ }\textbf {\bibinfo {volume} {39}},\ \bibinfo
  {pages} {2740} (\bibinfo {year} {1989})}\BibitemShut {NoStop}%
\bibitem [{\citenamefont {Ruhman}\ and\ \citenamefont
  {Lee}(2017)}]{Ruhman2017}%
  \BibitemOpen
  \bibfield  {author} {\bibinfo {author} {\bibfnamefont {J.}~\bibnamefont
  {Ruhman}}\ and\ \bibinfo {author} {\bibfnamefont {P.~A.}\ \bibnamefont
  {Lee}},\ }\bibfield  {title} {\bibinfo {title} {Pairing from dynamically
  screened coulomb repulsion in bismuth},\ }\href
  {https://doi.org/10.1103/PhysRevB.96.235107} {\bibfield  {journal} {\bibinfo
  {journal} {Phys. Rev. B}\ }\textbf {\bibinfo {volume} {96}},\ \bibinfo
  {pages} {235107} (\bibinfo {year} {2017})}\BibitemShut {NoStop}%
\bibitem [{\citenamefont {Fatemi}\ and\ \citenamefont
  {Ruhman}(2018)}]{Fatemi2018}%
  \BibitemOpen
  \bibfield  {author} {\bibinfo {author} {\bibfnamefont {V.}~\bibnamefont
  {Fatemi}}\ and\ \bibinfo {author} {\bibfnamefont {J.}~\bibnamefont
  {Ruhman}},\ }\bibfield  {title} {\bibinfo {title} {Synthesizing coulombic
  superconductivity in van der waals bilayers},\ }\href
  {https://doi.org/10.1103/PhysRevB.98.094517} {\bibfield  {journal} {\bibinfo
  {journal} {Phys. Rev. B}\ }\textbf {\bibinfo {volume} {98}},\ \bibinfo
  {pages} {094517} (\bibinfo {year} {2018})}\BibitemShut {NoStop}%
\bibitem [{\citenamefont {Ruvalds}(1981)}]{Ruvalds1981}%
  \BibitemOpen
  \bibfield  {author} {\bibinfo {author} {\bibfnamefont {J.}~\bibnamefont
  {Ruvalds}},\ }\bibfield  {title} {\bibinfo {title} {Are there acoustic
  plasmons?},\ }\href {https://doi.org/10.1080/00018738100101427} {\bibfield
  {journal} {\bibinfo  {journal} {Advances in Physics}\ }\textbf {\bibinfo
  {volume} {30}},\ \bibinfo {pages} {677} (\bibinfo {year} {1981})},\ \Eprint
  {https://arxiv.org/abs/https://doi.org/10.1080/00018738100101427}
  {https://doi.org/10.1080/00018738100101427} \BibitemShut {NoStop}%
\bibitem [{\citenamefont {Svintsov}\ and\ \citenamefont
  {Ryzhii}(2019)}]{Svintsov2019_PRL}%
  \BibitemOpen
  \bibfield  {author} {\bibinfo {author} {\bibfnamefont {D.}~\bibnamefont
  {Svintsov}}\ and\ \bibinfo {author} {\bibfnamefont {V.}~\bibnamefont
  {Ryzhii}},\ }\bibfield  {title} {\bibinfo {title} {Comment on ``{N}egative
  {L}andau {D}amping in {B}ilayer {G}raphene''},\ }\href
  {https://doi.org/10.1103/PhysRevLett.123.219401} {\bibfield  {journal}
  {\bibinfo  {journal} {Phys. Rev. Lett.}\ }\textbf {\bibinfo {volume} {123}},\
  \bibinfo {pages} {219401} (\bibinfo {year} {2019})}\BibitemShut {NoStop}%
\bibitem [{\citenamefont {Svintsov}(2019)}]{Svintsov2019_PRB}%
  \BibitemOpen
  \bibfield  {author} {\bibinfo {author} {\bibfnamefont {D.}~\bibnamefont
  {Svintsov}},\ }\bibfield  {title} {\bibinfo {title} {Emission of plasmons by
  drifting {D}irac electrons: {A} hallmark of hydrodynamic transport},\ }\href
  {https://doi.org/10.1103/PhysRevB.100.195428} {\bibfield  {journal} {\bibinfo
   {journal} {Phys. Rev. B}\ }\textbf {\bibinfo {volume} {100}},\ \bibinfo
  {pages} {195428} (\bibinfo {year} {2019})}\BibitemShut {NoStop}%
\bibitem [{\citenamefont {Pines}\ and\ \citenamefont
  {Schrieffer}(1962)}]{PinesSchrieffer1962}%
  \BibitemOpen
  \bibfield  {author} {\bibinfo {author} {\bibfnamefont {D.}~\bibnamefont
  {Pines}}\ and\ \bibinfo {author} {\bibfnamefont {J.~R.}\ \bibnamefont
  {Schrieffer}},\ }\bibfield  {title} {\bibinfo {title} {Approach to
  {E}quilibrium of {E}lectrons, {P}lasmons, and {P}honons in {Q}uantum and
  {C}lassical {P}lasmas},\ }\href {https://doi.org/10.1103/PhysRev.125.804}
  {\bibfield  {journal} {\bibinfo  {journal} {Phys. Rev.}\ }\textbf {\bibinfo
  {volume} {125}},\ \bibinfo {pages} {804} (\bibinfo {year}
  {1962})}\BibitemShut {NoStop}%
\bibitem [{\citenamefont {Hofmann}\ \emph {et~al.}(2015)\citenamefont
  {Hofmann}, \citenamefont {Barnes},\ and\ \citenamefont
  {Das~Sarma}}]{Hofmann2015_Liquid}%
  \BibitemOpen
  \bibfield  {author} {\bibinfo {author} {\bibfnamefont {J.}~\bibnamefont
  {Hofmann}}, \bibinfo {author} {\bibfnamefont {E.}~\bibnamefont {Barnes}},\
  and\ \bibinfo {author} {\bibfnamefont {S.}~\bibnamefont {Das~Sarma}},\
  }\bibfield  {title} {\bibinfo {title} {Interacting {D}irac liquid in
  three-dimensional semimetals},\ }\href
  {https://doi.org/10.1103/PhysRevB.92.045104} {\bibfield  {journal} {\bibinfo
  {journal} {Phys. Rev. B}\ }\textbf {\bibinfo {volume} {92}},\ \bibinfo
  {pages} {045104} (\bibinfo {year} {2015})}\BibitemShut {NoStop}%
\bibitem [{\citenamefont {Principi}\ \emph {et~al.}(2011)\citenamefont
  {Principi}, \citenamefont {Asgari},\ and\ \citenamefont
  {Polini}}]{Principi2011}%
  \BibitemOpen
  \bibfield  {author} {\bibinfo {author} {\bibfnamefont {A.}~\bibnamefont
  {Principi}}, \bibinfo {author} {\bibfnamefont {R.}~\bibnamefont {Asgari}},\
  and\ \bibinfo {author} {\bibfnamefont {M.}~\bibnamefont {Polini}},\
  }\bibfield  {title} {\bibinfo {title} {Acoustic plasmons and composite
  hole-acoustic plasmon satellite bands in graphene on a metal gate},\ }\href
  {https://doi.org/https://doi.org/10.1016/j.ssc.2011.07.015} {\bibfield
  {journal} {\bibinfo  {journal} {Solid State Communications}\ }\textbf
  {\bibinfo {volume} {151}},\ \bibinfo {pages} {1627 } (\bibinfo {year}
  {2011})}\BibitemShut {NoStop}%
\end{thebibliography}%

\end{document}